\documentclass[reprint,prd,tightenlines,floatfix,
nofootinbib,eqsecnum,superscriptaddress,onecolumn]{revtex4-2}

\usepackage[T1]{fontenc}		

\usepackage{amsmath,amsfonts,amssymb,amstext,mathrsfs}
\usepackage{mathpazo}

\usepackage[dvips]{graphicx}
\usepackage{epsf,float}
\usepackage{revsymb}

\usepackage{dcolumn}
\usepackage{braket}
\usepackage{color,xcolor}
\usepackage{graphicx}
\usepackage{subfigure}
\usepackage{multirow}
\usepackage{tabularx}
\usepackage{pstricks}
\usepackage[section]{placeins}
\usepackage{booktabs}
\usepackage{array}

\usepackage{hyperref}

\usepackage{lineno}

\newcommand{\Pom}{\mathbb{P}}

\newcommand{\Reg}{\mathbb{R}}

\newcommand{\bdPt}{\mbox{\boldmath $dP_{t}$}}
\newcommand{\bqta}{\mbox{\boldmath $q_{t,1}$}}
\newcommand{\bqtb}{\mbox{\boldmath $q_{t,2}$}}
\newcommand{\bpta}{\mbox{\boldmath $p_{t,1}$}}
\newcommand{\bptb}{\mbox{\boldmath $p_{t,2}$}}

\newcommand{\bptat}{\mbox{\boldmath $\tilde{p}_{t,1}$}}
\newcommand{\bptbt}{\mbox{\boldmath $\tilde{p}_{t,2}$}}
\newcommand{\bkt}{\mbox{\boldmath $l_{t}$}}
\newcommand{\bktsqrt}{\mbox{\boldmath{$l_{t}^{2}$}}}

\newcommand{\bk}{\mbox{\boldmath $k$}}

\renewcommand\slash[1]{\not \! #1}

\newcommand{\p}{\partial}

\newcommand{\twosidep}[1]{\stackrel{\leftrightarrow}{\p}_{\! #1}}

\usepackage[normalem]{ulem}  

\begin{document}

\title{Central exclusive production of $\eta$ and $\eta'$ mesons in diffractive proton-proton collisions \\
at the LHC within the tensor-pomeron approach}

\author{Piotr Lebiedowicz}
\email{Piotr.Lebiedowicz@ifj.edu.pl}
\affiliation{Institute of Nuclear Physics Polish Academy of Sciences, Radzikowskiego 152, PL-31342 Krak{\'o}w, Poland}

\author{Otto Nachtmann}
\email{O.Nachtmann@thphys.uni-heidelberg.de}
\affiliation{Institut f\"ur Theoretische Physik, Universit\"at Heidelberg, Philosophenweg 16, D-69120 Heidelberg, Germany}

\author{Antoni Szczurek}
\email{Antoni.Szczurek@ifj.edu.pl}
\affiliation{Institute of Nuclear Physics Polish Academy of Sciences, Radzikowskiego 152, PL-31342 Krak{\'o}w, Poland}
\affiliation{Institute of Physics, Faculty of Exact and Technical Sciences, University of Rzesz{\'o}w, 
Pigonia 1, PL-35310 Rzesz{\'o}w, Poland}

\begin{abstract}
We present a study of the central exclusive production (CEP)
of $\eta$ and $\eta'(958)$ mesons 
in diffractive proton-proton collisions at high energies.
The amplitudes, including pomeron and $f_{2 \Reg}$ reggeon exchanges, 
are calculated within the tensor-pomeron model. 
Absorption effects are also taken into account
at the amplitude level.
We fit some undetermined model parameters 
(coupling constants and cutoff parameters in form factors)
to the WA102 experimental data 
and then make predictions for the LHC energy $\sqrt{s} = 13$~TeV.
Both, total cross sections and several differential distributions are presented.
For $pp \to pp \eta$, we find an upper limit
for the total cross section 
of 2.5~$\mu$b for pseudorapidity of the $\eta$ meson $|\eta_{M}| < 1$
and 5.6~$\mu$b for $2 < \eta_{M} < 5$.
For $pp \to pp \eta'$, we predict the cross section
to be in the range of
0.3--0.7~$\mu$b for pseudorapidity of the $\eta'$ meson $|\eta_{M}| < 1$
and
0.9--2.1~$\mu$b for $2 < \eta_{M} < 5$.
This opens the possibility to study diffractive production of pseudoscalar mesons in experiments at the LHC.
We discuss if there are arguments from SU(3)-flavor symmetry
which would forbid pomeron-pomeron fusion giving an $\eta$ meson.
In our opinion such arguments do not exist.
We also consider CEP of the pseudoscalars $\eta$ and $\eta'(958)$
and the pseudovector meson $f_{1}(1285)$ in diffractive
proton-proton collisions
in a theory with a scalar pomeron. 
We show that none of these particles can be produced in this way
in the scalar-pomeron theory. Thus, experimental observation
of any of these particles in the above CEP processes at the LHC 
would give striking evidence against a scalar character of the pomeron.
\end{abstract}

\maketitle

\section{Introduction}
\label{sec:1}

Central exclusive production (CEP) of $\eta$ and $\eta'$ mesons
with $J^{PC} = 0^{-+}$
in proton-proton collisions 
has been a subject of both theoretical and experimental studies.
Diffractive production of pseudoscalar mesons at high energies
can be mediated by the double-pomeron-exchange mechanism
\cite{Lebiedowicz:2013ika}.
The Born-level amplitudes for $\eta$ and $\eta'$ production were derived in \cite{Lebiedowicz:2013ika}
in the tensor-pomeron approach \cite{Ewerz:2013kda}.
The effective coupling vertices for the fusion of two tensor pomerons 
into a pseudoscalar meson
were derived from the corresponding interaction Lagrangians.
Within this framework, there are 
two independent $\Pom \Pom \eta$ and $\Pom \Pom \eta'$ couplings possible,
corresponding to the allowed values of $(l, S) = (1,1)$ and $(3,3)$,
where $l$ denotes the orbital angular momentum 
and $S$ the total spin of the two tensor pomerons.
To be precise, $l$ and $S$ refer to the fictitious reaction
of two ``real'' spin~2 pomerons fusing to give $\eta$ or $\eta'$;
see (A.10)--(A.16) and Table~6 of \cite{Lebiedowicz:2013ika}.
The covariant couplings of the pomerons to these particles
are then constructed accordingly.
The corresponding coupling constants 
were fitted to differential distributions 
of the WA102 Collaboration \cite{WA102:1998ixr}
at $\sqrt{s} = 29.1$~GeV
and to the total cross sections 
given in Table~1 of \cite{Kirk:2000ws}.
We cannot exclude the possibility that secondary reggeon exchanges
may play an important role in the WA102 energy range;
see Sec.~3.2 of \cite{Lebiedowicz:2013ika} 
and the discussion in \cite{Lebiedowicz:2020yre}.
It is found in \cite{Lebiedowicz:2013ika}
that for the $\eta$ meson production the inclusion of additional tensorial contributions
of $f_{2 \Reg} \Pom$, $\Pom f_{2 \Reg}$, and $f_{2 \Reg} f_{2 \Reg}$
to the $\Pom \Pom$ contribution improves
the description of experimental differential distributions.
Production of $\eta'$ meson seems to be less affected 
by contributions from subleading exchanges.

The purpose of our analysis is to provide theoretical predictions 
for the production of $\eta$ and $\eta'$
in the $pp \to pp \eta(\eta')$ reactions
within the tensor-pomeron approach 
for experiments at the LHC, 
where pomeron-pomeron fusion is the dominant production mechanism.
In the present work we improve the results given in \cite{Lebiedowicz:2013ika} and extend them to LHC energies.
The consideration of adding absorption effects 
to the Born amplitudes
due to proton-proton interactions
is an important point of the current analysis.
Some of the model parameters can then be estimated 
by comparing the model results with the WA102 data, 
as was done for the CEP of $f_{1}$ mesons in \cite{Lebiedowicz:2020yre}.

It is worth noting that CEP of $\eta$ and $\eta'$ mesons 
was discussed earlier in 
\cite{Kochelev:1999wv,Kochelev:2000wm,Shuryak:2003xz,Petrov:2004hh,Ryutin:2014eua}.
We believe, however, that pomeron couplings that are essential 
in the construction of matrix elements for a given type of soft process
should be treated as tensor couplings.
Some remarks on different views of
the pomeron were made in \cite{Ewerz:2016onn}. 
It was shown there that only the tensor ansatz 
for the soft pomeron is compatible 
with the STAR results from polarized 
elastic proton-proton scattering \cite{Adamczyk:2012kn}.
In \cite{Britzger:2019lvc}, the authors gave further strong
evidence against the hypothesis that the pomeron 
has vector character.
In light of the arguments in these works,
we cannot support 
the conclusions of \cite{Close:1999is,Close:1999bi}
that the pomeron 
transforms as a non-conserved vector current
as was deduced from the comparison of their model results with data
measured by the WA102 Collaboration \cite{WA102:1999poj}.
In our articles, see for examples
Refs.~\cite{Lebiedowicz:2013ika,Lebiedowicz:2016ioh,Lebiedowicz:2018eui,Lebiedowicz:2019jru,Lebiedowicz:2019por,Lebiedowicz:2019boz,Lebiedowicz:2020yre,Lebiedowicz:2021pzd},
we showed that the tensor-pomeron concept 
for the central production of mesons
works quite well in reproducing the data when available.

In \cite{Anderson:2014jia,Anderson:2016zon} the $pp \to pp \eta$ process
was discussed using the Sakai-Sugimoto model.
There, also a comparison of the model with the total cross section
measured by the WA102 Collaboration \cite{WA102:1998ixr} 
was done.
The total cross section calculated in the Sakai-Sugimoto model 
was found to be about an order of magnitude smaller than the WA102 result.
For some related analyses based on the holographic model of QCD 
see \cite{Hechenberger:thesis}.
Finally, we note that in \cite{Lebiedowicz:2020yre}
the relation between two different approaches,
the tensor-pomeron model and the Sakai-Sugimoto model,
for the $\Pom \Pom f_{1}$ couplings was discussed.
By freely adjusting the coupling constants in both models, considering only the pomeron-pomeron fusion mechanism, 
a reasonable agreement with the WA102 data on CEP of $f_{1}$ was obtained.
Due to the potential impact of secondary reggeon exchanges at lower energies, where the parameters of our model were fixed, our predictions for the LHC given there should be considered as upper limits of the cross sections.\footnote{If at the WA102 energies there are important contributions
from subleading reggeon-pomeron and reggeon-reggeon fusion terms,
the cross sections at LHC energies could be significantly smaller.
The reduction could be up to a factor of 4 as estimated in Appendix~D of \cite{Lebiedowicz:2020yre} for the CEP of $f_{1}(1285)$.}

Experimental studies of processes involving light mesons 
in the final state at LHC energies
would be particularly helpful in order to get a better understanding
of various theoretical aspects of these reactions.
We will learn, for instance, about pomeron-pomeron-meson couplings, absorption effects, and the role of subleading exchanges.
Of particular interest is the transition between 
the nonperturbative (small meson transverse momenta) 
and perturbative (large meson transverse momenta) regimes. 
The perturbative mechanism for CEP of pseudoscalar mesons
was considered in \cite{Szczurek:2006bn,Harland-Lang:2013ncy}.

Our paper is organized as follows.
In the next section we discuss first the scalar-pomeron case.
Then we give analytic expressions
for the amplitudes for the $pp \to pp \eta(\eta')$ reactions
in the tensor-pomeron model.
The results of our calculations
are presented in Sec.~\ref{sec:3}.
Section~\ref{sec:4} contains our conclusions.
In Appendix~\ref{sec:A1} we discuss 
the $\Pom f_{2 \Reg} \eta(\eta')$ couplings
needed for our calculations.

\section{Diffractive production of $\eta$ and $\eta'(958)$ mesons in proton-proton collisions}
\label{sec:2}
We study central exclusive production of $\eta$ and $\eta'(958)$
in proton-proton collisions 
\begin{eqnarray}
p(p_{a},\lambda_{a}) + p(p_{b},\lambda_{b}) \to
p(p_{1},\lambda_{1}) + M(k) + p(p_{2},\lambda_{2}) \,,
\label{2to3_reaction}
\end{eqnarray}
where $p_{a,b}$, $p_{1,2}$ and $\lambda_{a,b}$, 
$\lambda_{1,2} = \pm \frac{1}{2}$
denote the four-momenta and helicities of the protons, respectively.
The four-momentum of the meson $M$ ($\eta$ or $\eta'(958)$)
with $I^{G}(J^{PC}) = 0^{+}(0^{-+})$ is denoted by $k$.
\begin{figure}[!ht]
\includegraphics[width=6.cm]{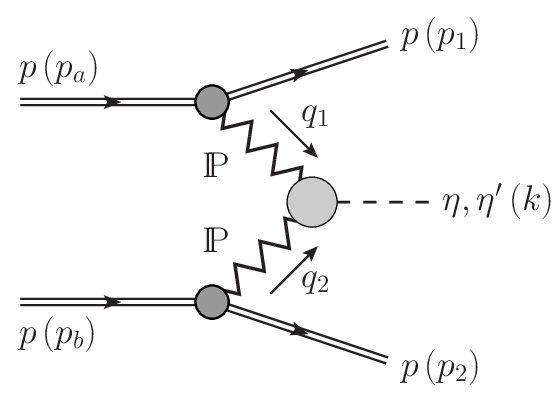}
\caption{\label{fig:diagram}
\small
The Born-level diagram for the reaction (\ref{2to3_reaction})
with double-pomeron exchange (i.e., $\Pom \Pom$-fusion mechanism).}
\end{figure}

We focus here on the $pp \to pp M$ processes
mediated by double-pomeron exchange,
shown at the Born level by the diagram in Fig.~\ref{fig:diagram}.
Let us first assume that the pomeron has a scalar character
replacing $\Pom$ by a scalar $\Pom_{\rm S}$ in Fig.~\ref{fig:diagram}.
We are then dealing with the vertex function
$\Gamma^{(\Pom_{\rm S} \Pom_{\rm S} \to M)}$ which
can only be a function of $q_{1}^{2}$, $q_{2}^{2}$, and $k^{2}$,
due to Lorentz invariance:
\begin{align}
\includegraphics[width=.20\textwidth]{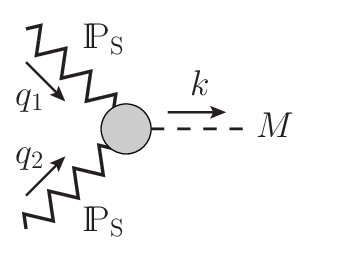}
i \Gamma^{(\Pom_{\rm S} \Pom_{\rm S} \to M)}(q_{1}^{2},q_{2}^{2},k^{2})\,.
\label{2.1a}
\end{align}
In strong interactions we have parity invariance.
For a pseudoscalar meson $M$ we must have
\begin{eqnarray}
\Gamma^{(\Pom_{\rm S} \Pom_{\rm S} \to M)}(q_{1}'^{2},q_{2}'^{2},k'^{2})
=
-\Gamma^{(\Pom_{\rm S} \Pom_{\rm S} \to M)}(q_{1}^{2},q_{2}^{2},k^{2})\,,
\label{2.1b}
\end{eqnarray}
where
\begin{eqnarray}
{q_{1}'}^{\mu} &=& {\cal P}^{\mu}_{\ \  \nu} \ q_{1}^{\nu}\,, \qquad
{q_{2}'}^{\mu} = {\cal P}^{\mu}_{\ \  \nu} \ q_{2}^{\nu}\,,\qquad
{k'}^{\mu} = {\cal P}^{\mu}_{\ \  \nu} \ k^{\nu}\,, \nonumber \\
({\cal P}^{\mu}_{\ \  \nu}) &=& {\rm diag}(1,-1,-1,-1)\,.
\label{2.1c}
\end{eqnarray}
But we have
\begin{eqnarray}
q_{1}'^{2} = q_{1}^{2}\,,  \qquad
q_{2}'^{2} = q_{2}^{2}\,,  \qquad
k'^{2} = k^{2}\,.
\label{2.1d}
\end{eqnarray}
Therefore, we conclude:
\begin{eqnarray}
\Gamma^{(\Pom_{\rm S} \Pom_{\rm S} \to M)}(q_{1}^{2},q_{2}^{2},k^{2}) = 0 \,.
\label{2.1e}
\end{eqnarray}
With a scalar pomeron CEP of a pseudoscalar meson $M$ with
$I^{G}(J^{PC}) = 0^{+}(0^{-+})$ is not possible.

In a similar way we discuss the coupling of 
an axial-vector meson $\widetilde{M}$
where $I^{G}(J^{PC}) = 0^{+}(1^{++})$
to two scalar pomerons
\begin{align}
\includegraphics[width=.20\textwidth]{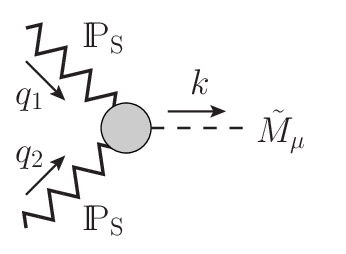}
i \Gamma_{\mu}^{(\Pom_{\rm S} \Pom_{\rm S} \to \widetilde{M})}(q_{1},q_{2})\,.
\end{align}
Here Lorentz invariance requires
\begin{align}
\Gamma^{\mu\,(\Pom_{\rm S} \Pom_{\rm S} \to \widetilde{M})}(q_{1},q_{2})=
(q_{1} + q_{2})^{\mu} \, A(q_{1}^{2},q_{2}^{2},k^{2}) + 
(q_{1} - q_{2})^{\mu} \, B(q_{1}^{2},q_{2}^{2},k^{2}) 
\,,
\label{2.1g}
\end{align}
where $A(.)$ and $B(.)$ are invariant functions.
Requiring in (\ref{2.1g}) parity invariance leads to
\begin{align}
\Gamma^{\mu\,(\Pom_{\rm S} \Pom_{\rm S} \to \widetilde{M})}(q'_{1},q'_{2}) =
- {\cal P}^{\mu}_{\ \  \nu}
\Gamma^{\nu\,(\Pom_{\rm S} \Pom_{\rm S} \to \widetilde{M})}(q_{1},q_{2})\,,
\label{2.1h}
\end{align}
which implies
\begin{align}
A(q_{1}^{2},q_{2}^{2},k^{2}) =
B(q_{1}^{2},q_{2}^{2},k^{2}) = 0\,.
\label{2.1i}
\end{align}
CEP of an axial-vector meson $\widetilde{M}$
is not possible with a scalar pomeron.
A prominent axial-vector meson is the $f_{1}(1285)$.

In our paper
we treat the reactions (\ref{2to3_reaction})
in the tensor-pomeron approach
as introduced in \cite{Ewerz:2013kda}.
The pomeron ($\Pom$) with the charge conjugation $C = +1$ 
is described as effective rank-two symmetric tensor exchange.

The kinematic variables are
\begin{eqnarray}
&&q_1 = p_{a} - p_{1}, \quad q_2 = p_{b} - p_{2}, \quad k = q_{1} + q_{2}, \nonumber \\
&&t_1 = q_{1}^{2}, \quad t_2 = q_{2}^{2}, \nonumber \\
&&u_1 = (p_{a} - k)^{2} = (p_{1} - q_{2})^{2}, \nonumber \\
&&u_2 = (p_{b} - k)^{2} = (p_{2} - q_{1})^{2}, \nonumber \\
&&s = (p_{a} + p_{b})^{2} = (p_{1} + p_{2} + k)^{2}, \nonumber \\
&&s_{1} = (p_{a} + q_{2})^{2} = (p_{1} + k)^{2}, \nonumber \\
&&s_{2} = (p_{b} + q_{1})^{2} = (p_{2} + k)^{2}, \nonumber \\
&&\nu_{1} = \frac{1}{4}(s_{1} - u_{1}) 
= \frac{1}{4}(p_{a}+p_{1}, q_{2}+k)
= \frac{1}{2}(p_{a}+p_{1}, k), \nonumber \\
&&\nu_{2} = \frac{1}{4}(s_{2} - u_{2}) 
= \frac{1}{4}(p_{b}+p_{2}, q_{1}+k)
= \frac{1}{2}(p_{b}+p_{2}, k)\,.
\label{2to3_kinematic}
\end{eqnarray}
For a detailed discussion of the kinematics see e.g. Appendix~D of \cite{Lebiedowicz:2013ika}.


The amplitude for the $\Pom \Pom$ fusion to 
a pseudoscalar particle or resonance $M$ reads
\begin{eqnarray}
{\cal {M}}_{pp \to pp M}^{(\Pom \Pom)} =
{\cal {M}}_{pp \to pp M}^{\rm Born} + 
{\cal {M}}_{pp \to pp M}^{pp-\rm{rescattering}}\,.
\label{amp_PP}
\end{eqnarray}
The second term represents the $pp$-rescattering corrections
to the first (Born) term.

The Born amplitude is given by
\begin{equation}
\begin{split}
{\cal M}^{\rm Born}_{\lambda_{a}\lambda_{b}\to\lambda_{1}\lambda_{2} M} 
= & (-i)\,
\bar{u}(p_{1}, \lambda_{1}) 
i\Gamma^{(\Pom pp)\,\mu_{1} \nu_{1}}(p_{1},p_{a}) 
u(p_{a}, \lambda_{a})\\
& \times 
i\Delta^{(\Pom)}_{\mu_{1} \nu_{1}, \alpha_{1} \beta_{1}}(2 \nu_{1},t_{1})\,
i\Gamma^{(\Pom \Pom M)\,\alpha_{1} \beta_{1},\alpha_{2} \beta_{2}}(q_{1},q_{2})\,
i\Delta^{(\Pom)}_{\alpha_{2} \beta_{2}, \mu_{2} \nu_{2}}(2 \nu_{2},t_{2}) \\
& \times
\bar{u}(p_{2}, \lambda_{2}) 
i\Gamma^{(\Pom pp)\,\mu_{2} \nu_{2}}(p_{2},p_{b}) 
u(p_{b}, \lambda_{b}) \,.
\end{split}
\label{amp_Born}
\end{equation}
The corresponding expressions
for the effective proton vertex function
and propagator for the tensor-pomeron exchange
are as follows (see Sec.~3 of \cite{Ewerz:2013kda})
\begin{eqnarray}
&&i\Gamma_{\mu \nu}^{(\Pom pp)}(p',p)
=-i 3 \beta_{\Pom NN} F_{1}(t)
\left\lbrace 
\frac{1}{2} 
\left[ \gamma_{\mu}(p'+p)_{\nu} 
     + \gamma_{\nu}(p'+p)_{\mu} \right]
- \frac{1}{4} g_{\mu \nu} (\slash{p}' + \slash{p})
\right\rbrace, \qquad
\label{add2}\\
&&i \Delta^{(\Pom)}_{\mu \nu, \kappa \lambda}(2 \nu_{1,2},t) = 
\frac{1}{8\nu_{1,2}} 
\left( g_{\mu \kappa} g_{\nu \lambda} 
     + g_{\mu \lambda} g_{\nu \kappa}
     - \frac{1}{2} g_{\mu \nu} g_{\kappa \lambda} \right)
(-i \,2 \nu_{1,2} \,\alpha'_{\Pom})^{\alpha_{\Pom}(t)-1}\,,
\label{add1}
\end{eqnarray}
where $t = (p'-p)^{2}$ and $\beta_{\Pom NN} = 1.87$~GeV$^{-1}$.
For simplicity we use for the pomeron-proton coupling 
the electromagnetic Dirac form factor $F_{1}(t)$ of the proton.
The pomeron trajectory $\alpha_{\Pom}(t)$ is
assumed to be of standard form
\begin{eqnarray}
\alpha_{\Pom}(t) = \alpha_{\Pom}(0)+\alpha'_{\Pom} t\,,
\quad 
\alpha_{\Pom}(0) = 1 + \epsilon_{\Pom} = 1.0808\,, 
\quad
\alpha'_{\Pom} = 0.25 \; \mathrm{GeV}^{-2}\,.
  \label{pomtrajectory}
\end{eqnarray}
In (\ref{amp_Born}) and (\ref{add1}) 
we use $2 \nu_{1}$ and $2 \nu_{2}$
instead of $s_{1}$ and $s_{2}$, respectively,
as variables in the pomeron propagators;
cf. (3.10) of \cite{Ewerz:2013kda}.
It is well known that in Regge theory $2 \nu$
is the variable which appears naturally because of its crossing properties;
see for instance chapter~6.4 of \cite{Collins:1977}
and chapter~6 of \cite{Ewerz:2013kda}.
At high energies and small $|t|$,
the kinematic regime considered in \cite{Ewerz:2013kda},
we have $2 \nu \to s$.
We have checked numerically that the use of $2 \nu_{1,2}$ instead of $s_{1,2}$
is significant at the WA102 energy ($\sqrt{s} = 29.1$~GeV) and leads to an increase of cross section of about 30--40\%, 
depending on the model parameters (i.e., on the share of exchange contributions) used.
At LHC energies, we can use with very good accuracy
the approximations $2 \nu_{1} \approx s_{1}$ 
and $2 \nu_{2} \approx s_{2}$,
especially for the midrapidity production of the mesons.

The $\Pom \Pom M$ coupling was discussed 
in Sec.~2.2 of \cite{Lebiedowicz:2013ika}.
The resulting $\Pom \Pom M$ vertex, including a form factor, 
is given as follows
[see (2.4) and (2.6) of \cite{Lebiedowicz:2013ika}]
\begin{eqnarray}
i\Gamma_{\mu \nu,\kappa \lambda}^{(\Pom \Pom M)} (q_{1},q_{2}) &=&
\left( 
  i\Gamma_{\mu \nu,\kappa \lambda}'^{(\Pom \Pom M)}(q_{1},q_{2}) \mid_{\rm bare}
+ i\Gamma_{\mu \nu,\kappa \lambda}''^{(\Pom \Pom M)}(q_{1},q_{2}) \mid_{\rm bare} 
\right)
F(q_{1}^{2},q_{2}^{2})\,,
\label{vertex_pompomPS}\\
i\Gamma_{\mu \nu,\kappa \lambda}'^{(\Pom \Pom M)}(q_{1}, q_{2})\mid_{\rm bare} &=&
i \, \frac{g_{\Pom \Pom M}'}{2 M_{0}} \,
\left( g_{\mu \kappa} \varepsilon_{\nu \lambda \rho \sigma}
      +g_{\nu \kappa} \varepsilon_{\mu \lambda \rho \sigma}
      +g_{\mu \lambda}\varepsilon_{\nu \kappa \rho \sigma}
      +g_{\nu \lambda}\varepsilon_{\mu \kappa \rho \sigma} \right) 
(q_{1}-q_{2})^{\rho} (q_{1}+q_{2})^{\sigma}\,,
\label{vertex_pompomPS_11}\\
i\Gamma_{\mu \nu,\kappa \lambda}''^{(\Pom \Pom M)}(q_{1}, q_{2})\mid_{\rm bare} &=&
i \, \frac{g_{\Pom \Pom M}''}{M_{0}^{3}} \, 
\bigg{\lbrace} \varepsilon_{\nu \lambda \rho \sigma} \left[ q_{1 \kappa} q_{2 \mu} - (q_{1} \cdot q_{2}) g_{\mu \kappa} \right] +
\varepsilon_{\mu \lambda \rho \sigma} \left[ q_{1 \kappa} q_{2 \nu} - (q_{1} \cdot q_{2}) g_{\nu \kappa} \right]  \nonumber \\
&&+
\varepsilon_{\nu \kappa \rho \sigma}  \left[ q_{1 \lambda} q_{2 \mu} - (q_{1} \cdot q_{2}) g_{\mu \lambda} \right] +
\varepsilon_{\mu \kappa \rho \sigma}  \left[ q_{1 \lambda} q_{2 \nu} - (q_{1} \cdot q_{2}) g_{\nu \lambda} \right] 
\bigg{\rbrace}
(q_{1}-q_{2})^{\rho} (q_{1}+q_{2})^{\sigma}\,. \qquad
\label{vertex_pompomPS_33}
\end{eqnarray}
In (\ref{vertex_pompomPS_11}) and (\ref{vertex_pompomPS_33})
$M_{0} \equiv 1$~GeV
and $g'_{\Pom \Pom M}$, $g''_{\Pom \Pom M}$ are dimensionless coupling constants that should be fitted to experimental data.
The $\Gamma_{\mu \nu,\kappa \lambda}'$ and 
$\Gamma_{\mu \nu,\kappa \lambda}''$ bare
vertices correspond to
$(l,S) = (1,1)$ and $(3,3)$, respectively,
as derived from the corresponding coupling Lagrangians
(2.3) and (2.5) in \cite{Lebiedowicz:2013ika}.

The form factor in the central vertex is parametrized as
\begin{eqnarray}
F(t_{1},t_{2}) = 
\frac{1}{1-t_{1}/\Lambda_{0}^{2}}
\frac{1}{1-t_{2}/\Lambda_{0}^{2}}\,,
\quad \Lambda_{0}^{2} = 0.5\;{\rm GeV}^{2}\,.
\label{Fpompommeson_FM}
\end{eqnarray}
%
%
%
Alternatively, we use the exponential form given by
\begin{eqnarray}
F(t_{1},t_{2}) = 
\exp\left( \frac{t_{1}+t_{2}}{\Lambda_{E}^{2}}\right) \,,
\label{Fpompommeson_exp}
\end{eqnarray}
where $\Lambda_{E}$ is the cutoff parameter fitted to the experimental data
as we will show below in Sec.~\ref{sec:3a}.

The amplitude representing absorption corrections 
due to the proton-proton interactions
can be written as
\begin{eqnarray}
{\cal M}_{pp \to pp M}^{pp-\rm{rescattering}}(s,\bpta,\bptb)=
\frac{i}{8 \pi^{2} s} \int d^{2}\bkt \,
{\cal M}_{pp \to pp}(s,-\bktsqrt)
{\cal M}_{pp\to pp M}^{\rm Born}
(s,\bptat,\bptbt)\,. \qquad \;\;
\label{amp_Abs}
\end{eqnarray}
Here, in the overall center-of-mass (c.m.) system, $\bpta$ and $\bptb$
are the transverse components of the momenta of the outgoing protons
and $\bkt$ is the transverse momentum carried around the pomeron loop.
${\cal M}_{pp\to pp M}^{\rm Born}$
is the Born amplitude given by (\ref{amp_Born})
with $\bptat = \bpta - \bkt$ and $\bptbt = \bptb + \bkt$.
${\cal M}_{pp \to pp}$
is the elastic $pp$ scattering amplitude given by (6.28)
in \cite{Ewerz:2013kda}
for large $s$ and with the momentum transfer $t=-\bktsqrt$.
In practice we work with the amplitudes in the high-energy approximation,
where $s$-channel-helicity conservation 
of the protons holds.

In addition to the $\Pom \Pom$ fusion, 
we also consider the $\Pom f_{2 \Reg}$, $f_{2 \Reg} \Pom$,
and $f_{2 \Reg} f_{2 \Reg}$ exchanges.
Since both $\Pom$ and $f_{2 \Reg}$ are described as tensor exchanges, 
the additional amplitudes, 
${\cal {M}}^{(\Pom f_{2 \Reg})}$, ${\cal {M}}^{(f_{2 \Reg} \Pom)}$, and ${\cal {M}}^{(f_{2 \Reg} f_{2 \Reg})}$,
are treated similarly to (\ref{amp_PP})--(\ref{vertex_pompomPS_33}).
The effective $f_{2 \Reg}$-proton vertex function
and the $f_{2 \Reg}$ propagator are given in \cite{Ewerz:2013kda}
by Eqs.~(3.49) and (3.12), respectively.
The important differences there are 
in the magnitude of the coupling constant 
and the parametrization of the trajectory:
\begin{eqnarray}
&&3 \beta_{\Pom NN} \to \frac{g_{f_{2 \Reg} pp}}{M_{0}}\,, \quad g_{f_{2 \Reg} pp} = 11.04\,,\\
&&\alpha_{\Pom}(t) \to \alpha_{\Reg_{+}}(t) =
\alpha_{\Reg_{+}}(0)+\alpha'_{\Reg_{+}}t\,, \quad
\alpha_{\Reg_{+}}(0) = 0.5475\,, \quad
\alpha'_{\Reg_{+}} = 0.9 \; \mathrm{GeV}^{-2}\,.
\label{amp_PR}
\end{eqnarray}
We use $2 \nu_{1,2}$ instead of $s_{1,2}$
as variables in the $f_{2 \Reg}$ propagators.

Note that in the reaction $\Pom f_{2 \Reg} \to M$ two \textit{different}
objects fuse to give the pseudoscalar meson $M$.
But with the \mbox{methods} of Appendix~A of \cite{Lebiedowicz:2013ika} 
we show in Appendix~\ref{sec:A1}
that also here we have only
the two couplings corresponding to $(l,S) = (1,1)$ and $(3,3)$.
Therefore, the bare vertices $\Pom f_{2 \Reg} M$
and $f_{2 \Reg} f_{2 \Reg} M$
are as the $\Pom \Pom M$ ones in (\ref{vertex_pompomPS_11}) 
and (\ref{vertex_pompomPS_33}), but with different coupling constants
$(g_{\Pom f_{2 \Reg} M}'\,, g_{\Pom f_{2 \Reg} M}'')$ and 
$(g_{f_{2 \Reg} f_{2 \Reg} M}'\,, g_{f_{2 \Reg} f_{2 \Reg} M}'')$,
respectively.
These coupling constants will be roughly fitted 
to existing central production data
from the WA102 experiment; see Table~\ref{tab:parameters} of Sec.~\ref{sec:3a} below.
We made the simplifying assumption 
that the form factor $F(t_{1},t_{2})$ 
is of the same type for all couplings.

\newpage
\section{Results}
\label{sec:3}

In this section we present our results for the reactions
$pp \to pp \eta$ and $pp \to pp \eta'$
at c.m. energy of \mbox{$\sqrt{s}=29.1$~GeV}
together with the WA102 experimental data
and our predictions for LHC experiments at $\sqrt{s}=13$~TeV.

\subsection{Comparison with the WA102 data}
\label{sec:3a}

According to Ref.~\cite{Kirk:2000ws}
the WA102 experimental cross sections 
for central production of $\eta$ and $\eta'(958)$ mesons
in $pp$ collisions at $\sqrt{s} = 29.1$~GeV are 
\begin{eqnarray}
\eta:  && \sigma_{\rm{exp.}} = (3859 \pm 368)~{\rm nb}\,,
\label{xs_eta}\\
\eta'(958): && \sigma_{\rm{exp.}} = (1717 \pm 184)~{\rm nb}\,.
\label{xs_etap}
\end{eqnarray}
In~\cite{WA102:1998ixr,Kirk:2000ws} also 
the distributions in $\phi_{pp}$, $t$, $\rm{dP_{t}}$, and $x_{F,M}$
for the $\eta$ and $\eta'$ production were presented.
Here, $t$ is the four-momentum transfer squared 
from one of the proton vertices
[we have $t = t_{1}$ or $t_{2}$; cf. (\ref{2to3_kinematic})], 
$\phi_{pp}$ is the azimuthal angle between 
the transverse momentum vectors $\bpta$ and $\bptb$ of the outgoing protons in the overall c.m. system,
${\rm dP_{t}}$ (the so-called ``glueball-filter variable'' 
\cite{Close:1997pj,Barberis:1996iq}) is defined as 
\begin{eqnarray}
\bdPt = \bqta - \bqtb = \bptb - \bpta \,, 
\quad {\rm dP_{t}} = |\bdPt|\,,
\label{dPt_variable}
\end{eqnarray}
and the Feynman-$x$ variable is defined as 
$x_{F,M} = 2 p_{z, M}/\sqrt{s}$
with $p_{z, M}$ the longitudinal momentum of the meson
in this c.m. system.

\begin{figure}[!ht]
\includegraphics[width=0.4\textwidth]{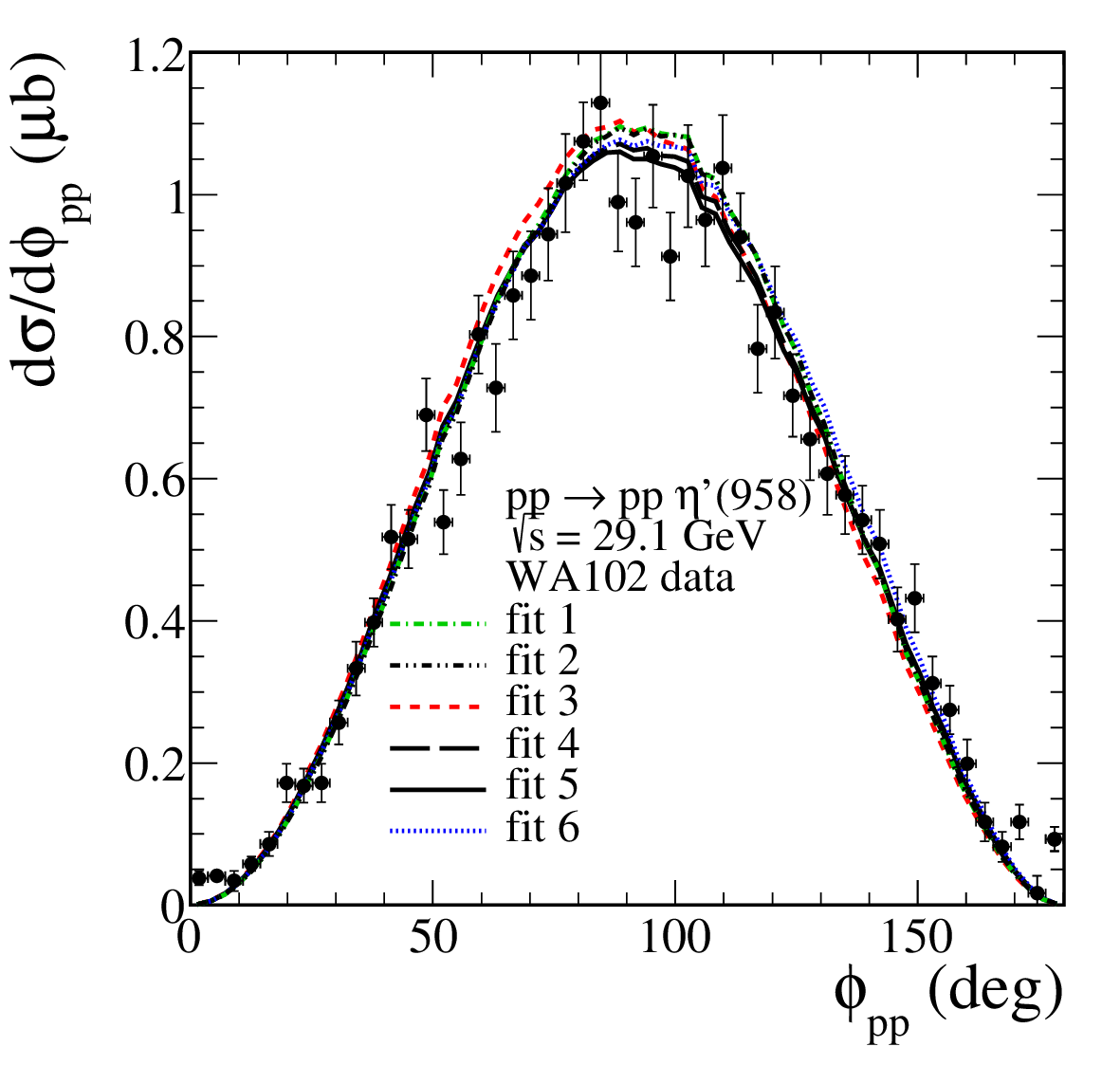}
\includegraphics[width=0.4\textwidth]{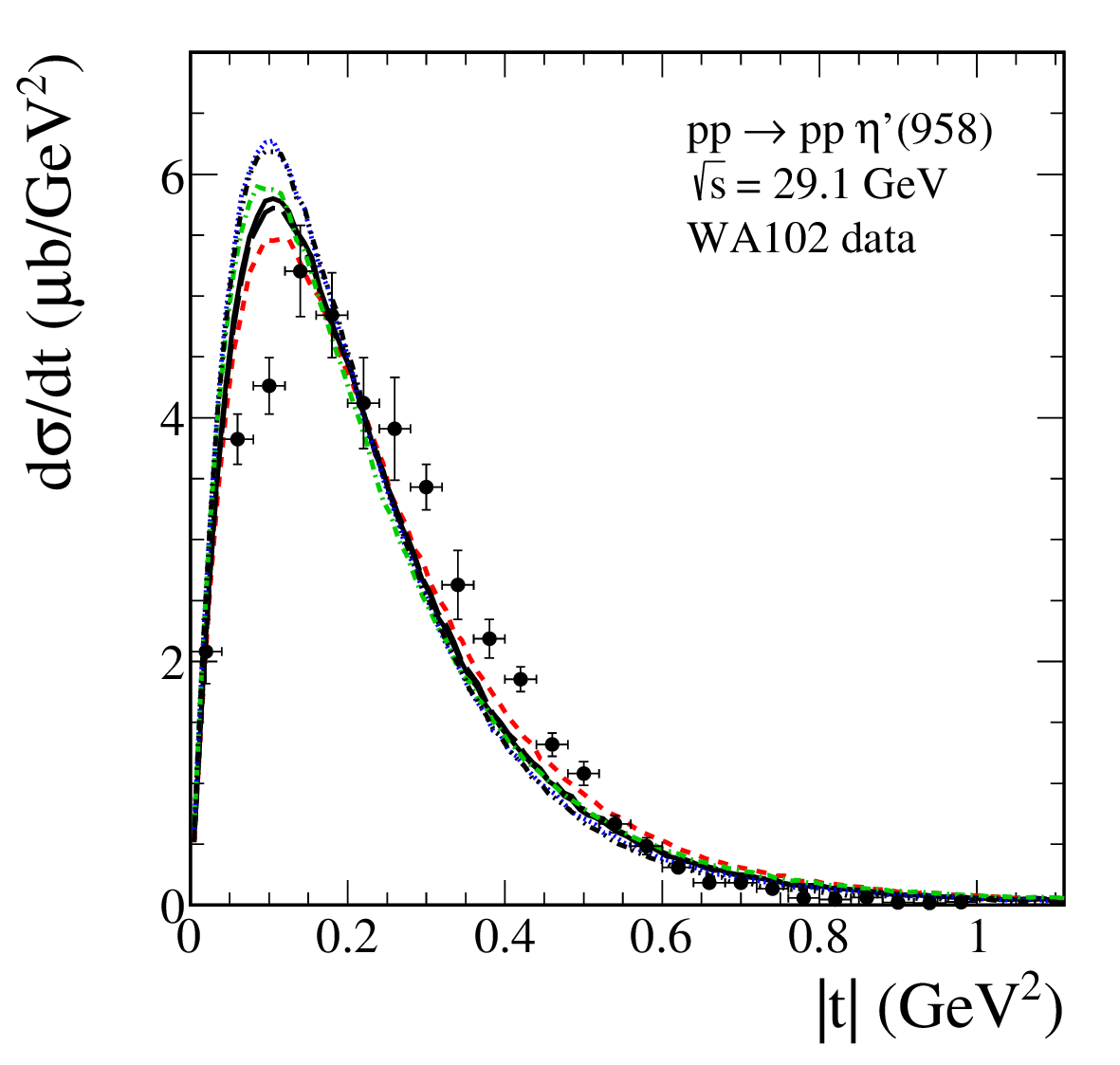}
\includegraphics[width=0.4\textwidth]{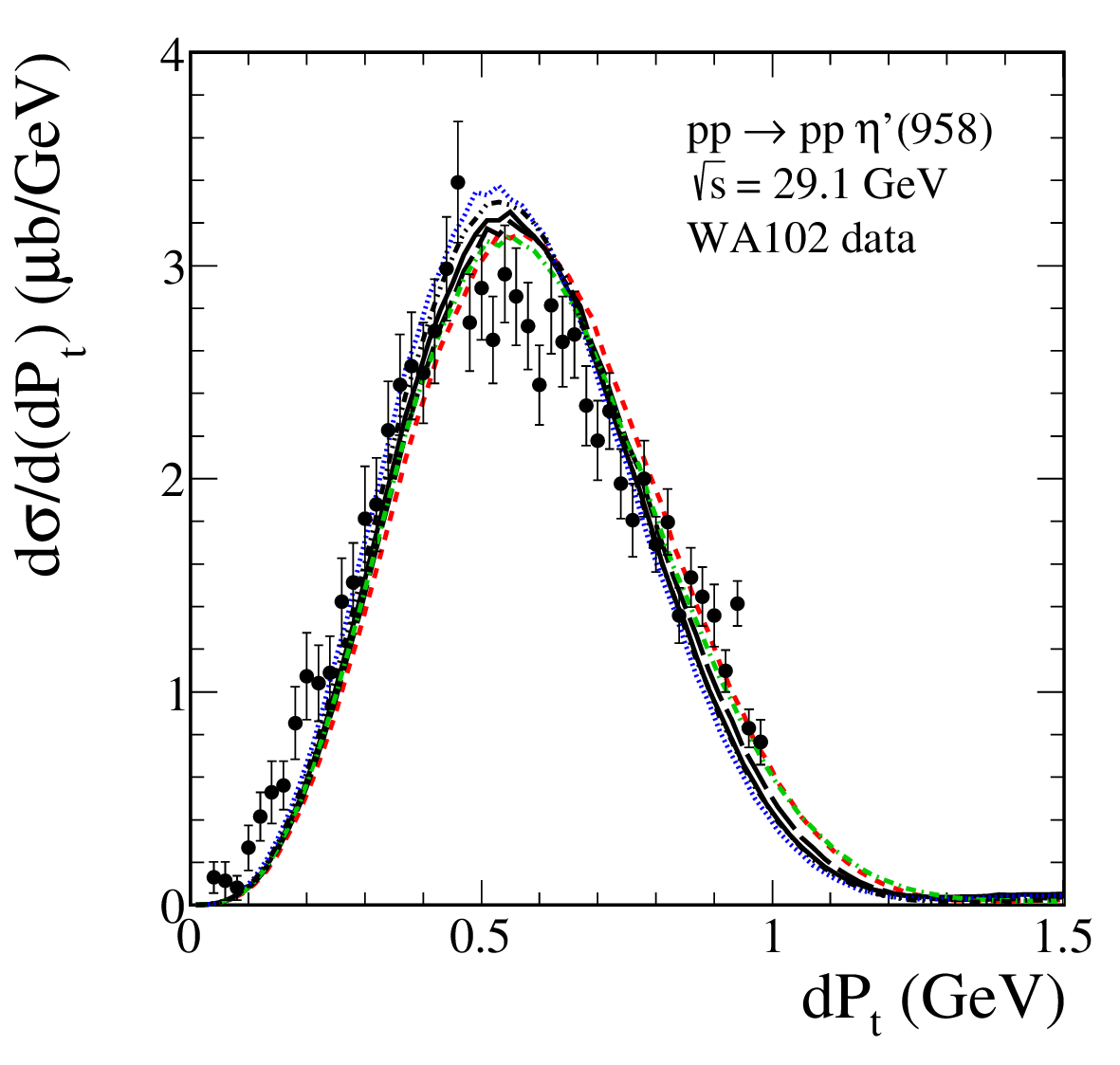}
\includegraphics[width=0.4\textwidth]{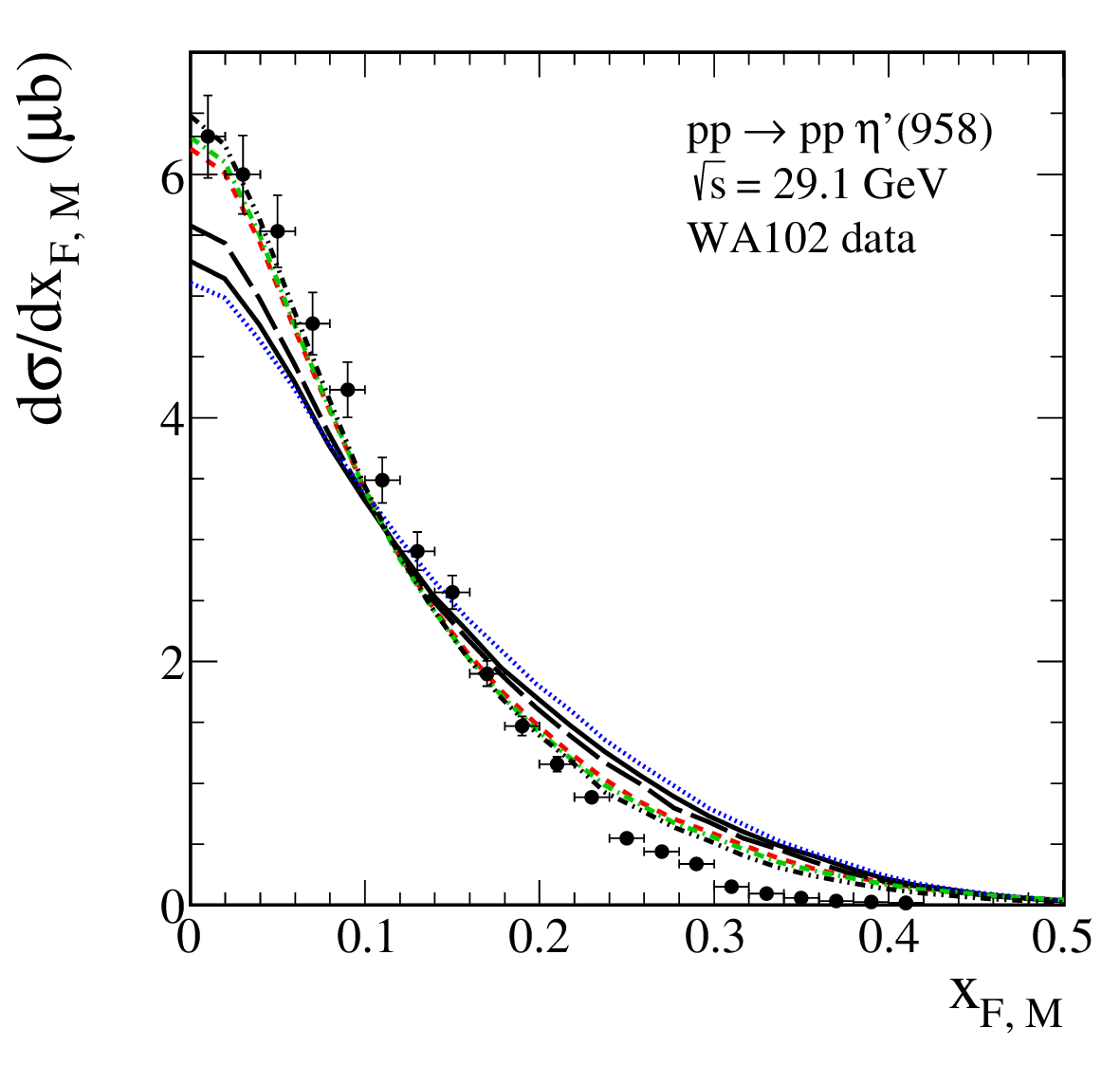}
  \caption{\label{fig:1}
  \small
Fits to the WA102 data \cite{WA102:1998ixr} 
for $\eta'$ meson production at $\sqrt{s} = 29.1$~GeV.
The absorption effects are included in the calculations.
For the line description see Table~\ref{tab:parameters}.}
\end{figure}

In Fig.~\ref{fig:1} we show the results for the $pp \to pp \eta'$ 
reaction at $\sqrt{s} = 29.1$~GeV.
The WA102 data points from \cite{WA102:1998ixr}
have been normalized to the mean value of the total cross section
(\ref{xs_etap}).
In Fits 1, 2, and 3 
we consider $\Pom \Pom$ plus $f_{2 \Reg} f_{2 \Reg}$ fusion processes.
By comparing the theoretical results 
and the WA102 distributions
we fixed the parameters 
in the $\Pom \Pom \eta'$ and $f_{2 \Reg} f_{2 \Reg} \eta'$ vertices.
We also present the results when 
the contribution of the exchanges 
$\Pom f_{2 \Reg}$ plus $f_{2 \Reg} \Pom$ 
are included in the calculation. 
These correspond to Fits~4, 5, and 6.
The numerical values of the parameters used in the calculations
are collected in Table~\ref{tab:parameters}.
We can see from Table~\ref{tab:parameters} how the choice of the type of the form factor $F(t_{1},t_{2})$ 
in (\ref{vertex_pompomPS})
and the cutoff parameter affect the strength of the different coupling constants.

\begin{table}[!h]
\caption{Parameters of the model determined
from the fit to the WA102 data at $\sqrt{s} = 29.1$~GeV
and the resulting cross sections 
with absorption effects are presented.
In the last column we present the ratio of the cross sections
with ($\sigma_{\rm abs}$) and without ($\sigma_{\rm Born}$) absorption effects, 
$S^{2} = \sigma_{\rm abs}/\sigma_{\rm Born}$.}
\label{tab:parameters}
\begin{tabular}{c|c|c|c|c|c|c|c|c|c|c}
\hline
\hline
Meson $M$ 
& Fit 
& $g'_{\Pom \Pom M}$ 
& $g''_{\Pom \Pom M}$ 
& $g'_{\Pom f_{2 \Reg} M}$ 
& $g''_{\Pom f_{2 \Reg} M}$ 
& $g'_{f_{2 \Reg} f_{2 \Reg} M}$ 
& $g''_{f_{2 \Reg} f_{2 \Reg} M}$ 
& Cutoff parameter
& $\sigma_{\rm abs}$ ($\mu$b) 
& $S^{2}$\\
\hline
$\eta'(958)$ 
& 1 & 2.7 & 1.2 & 0 & 0 & -10.0 & 0
&$\Lambda_{0}^{2} = 0.5\;{\rm GeV}^{2}$ 
& 1.74 & 0.60\\
& 2 & 2.7 & 1.2 & 0 & 0 & -10.0 & 0
&$\Lambda_{E} = 0.8\;{\rm GeV}$ 
& 1.74 & 0.63\\
& 3 & 2.05 & 1.2 & 0 & 0 & -10.0 & 0
&$\Lambda_{E} = 1.0\;{\rm GeV}$ 
& 1.75 & 0.60\\
& 4 & 1.6 & 1.2 & 1.8 & 0 & -10.0 & 0
&$\Lambda_{E} = 1.0\;{\rm GeV}$ 
& 1.72 & 0.62\\
& 5 & 1.4 & 1.4 & 2.3 & 0 & -10.0 & 0
&$\Lambda_{E} = 1.0\;{\rm GeV}$ 
& 1.75 & 0.65\\
& 6 & 1.4 & 1.4 & 2.5 & 0 & 10.0 & 0
&$\Lambda_{E} = 1.0\;{\rm GeV}$ 
& 1.75 & 0.63\\
& 7 & 1.4 & 1.4 & 1.3 & 0 & -25.0 & 0
&$\Lambda_{E} = 1.0\;{\rm GeV}$ 
& 1.75 & 0.66\\
& 8 & 1.4 & 1.4 & 2.15 & 0 & 25.0 & 0
&$\Lambda_{E} = 1.0\;{\rm GeV}$ 
& 1.74 & 0.65\\
\hline
$\eta$ 
& A & 1.8 & 1.8 & 0 & 0 & 10.0 & 0
&$\Lambda_{E} = 1.0\;{\rm GeV}$ 
& 3.73 & 0.66\\
& B & 1.0 & 1.0 & 1.1 & 3.0 & 4.0 & 0
&$\Lambda_{E} = 1.0\;{\rm GeV}$ 
& 3.55 & 0.73\\
& C & 1.0 & 1.0 & 1.1 & 3.0 & -4.0 & 0
&$\Lambda_{E} = 1.0\;{\rm GeV}$ 
& 3.64 & 0.67\\
& D & 1.0 & 1.0 & 1.7 & 4.0 & -4.0 & 0
&$\Lambda_{E} = 0.8\;{\rm GeV}$ 
& 3.66 & 0.73\\
\hline
\hline
\end{tabular}
\end{table}

For a deeper insight, we calculate the percentage share of different exchanges to the total cross section
for the $pp \to pp \eta'$ reaction at $\sqrt{s} = 29.1$~GeV.
In Fit~5, $\Pom \Pom$ accounts for 39\%, 
$\Pom f_{2 \Reg}$ + $f_{2 \Reg} \Pom$ for 28\%,
and $f_{2 \Reg} f_{2 \Reg}$ for only 3\%.
There is a large interference effect between these terms 
in the amplitude of about 30\% with respect to the total cross section.

For the CEP of the $\eta'$ meson,
the ratio of the cross sections 
measured by the WA102 experiment at $\sqrt{s} = 29.1$~GeV
to that measured by the WA76 experiment
at $\sqrt{s} = 12.7$~GeV 
was determined in \cite{WA102:1999poj} to be
\begin{eqnarray}
\frac{\sigma(\sqrt{s} = 29.1 \; {\rm GeV})}
{\sigma(\sqrt{s} = 12.7 \; {\rm GeV})}
=
0.72 \pm 0.16\,.
\label{ratio_exp}
\end{eqnarray}
The ratios we obtained for the parameter sets 1--6
from Table~\ref{tab:parameters}
are as follows
1.46, 1.47, 1.31, 1.29, 1.29, 1.19.
In the calculations, we assumed a rather large value of the coupling constant
$g'_{f_{2 \Reg} f_{2 \Reg} \eta'}$,
and despite this we do not reproduce the experimental result (\ref{ratio_exp}).
The discrepancy may indicate a certain role of other processes, e.g.
$\omega \omega$, $\rho^{0} \rho^{0}$, $\omega_{\Reg} \omega_{\Reg}$, 
and $\rho_{\Reg} \rho_{\Reg}$ exchanges.
To obtain a value of 0.7 for the ratio (\ref{ratio_exp}), 
we need $g'_{f_{2 \Reg} f_{2 \Reg} \eta'} \sim \pm 25.0$
as in sets 7 and 8 of Table~\ref{tab:parameters}.
Then, this parameter should be regarded as effective because it replaces the contribution of other (non-included) processes.
With the parameter set~7 and for $\sqrt{s} = 29.1$~GeV, 
the $\Pom \Pom$ contribution accounts for 39\%, 
$\Pom f_{2 \Reg}$ + $f_{2 \Reg} \Pom$ for 9\%,
and $f_{2 \Reg} f_{2 \Reg}$ for 22\% 
of the total cross section $\sigma(\sqrt{s} = 29.1~{\rm GeV}) = 1.75$~$\mu$b.
For $\sqrt{s} = 12.7$~GeV we obtain respectively
13\%,  6\%, and  61\% of 
$\sigma(\sqrt{s} = 12.7~{\rm GeV}) = 2.50$~$\mu$b.
As $\sqrt{s}$ increases from 12.7~GeV to 29.1~GeV, 
the cross section of the $f_{2 \Reg} f_{2 \Reg}$ exchange decreases 
by a factor of four, it increases slightly
for the $\Pom f_{2 \Reg} + f_{2 \Reg} \Pom$ fusion processes
and increases by a factor of 2.1 for the $\Pom \Pom$ fusion.

\begin{figure}[!ht]
\includegraphics[width=0.4\textwidth]{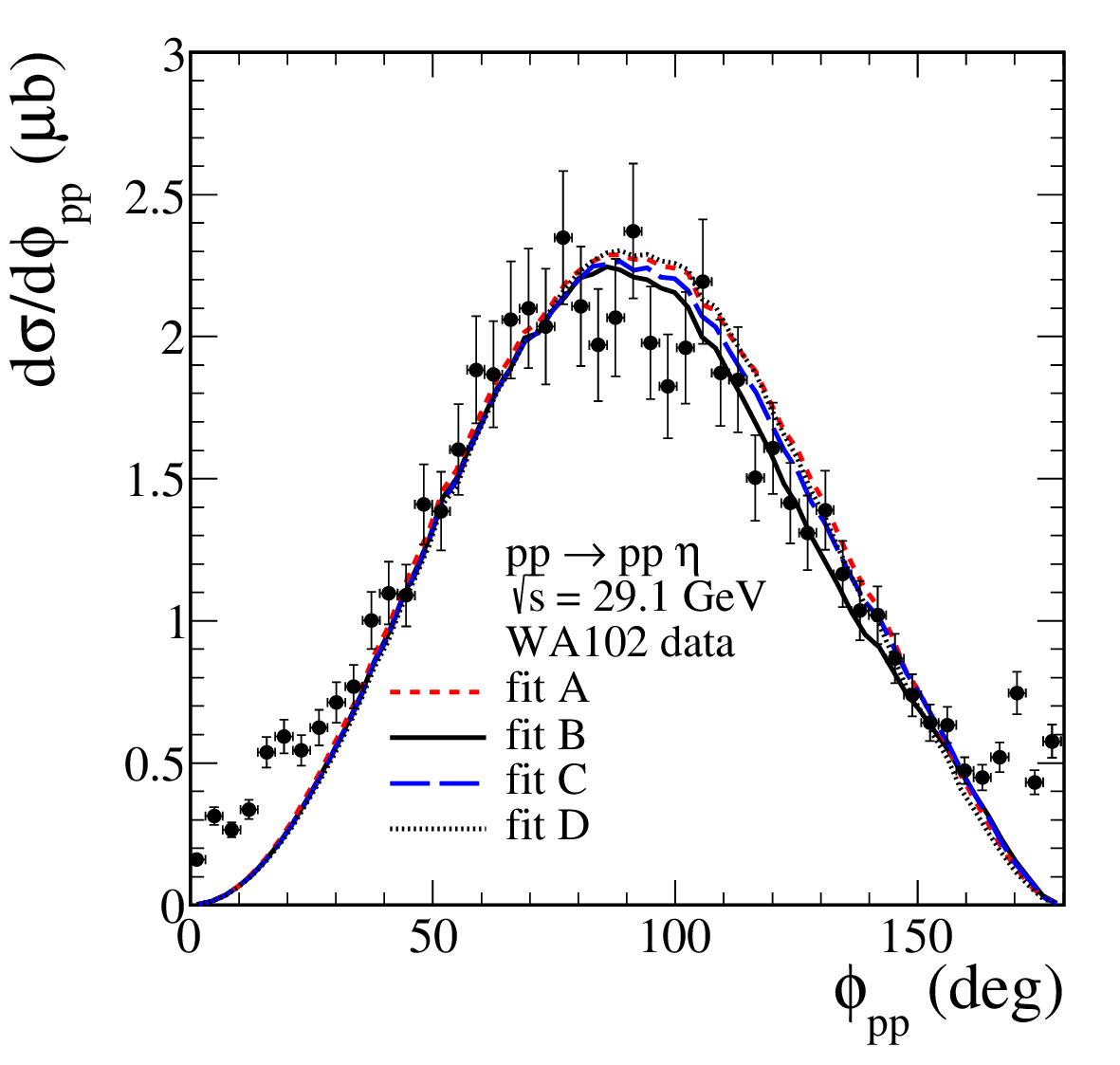}
\includegraphics[width=0.4\textwidth]{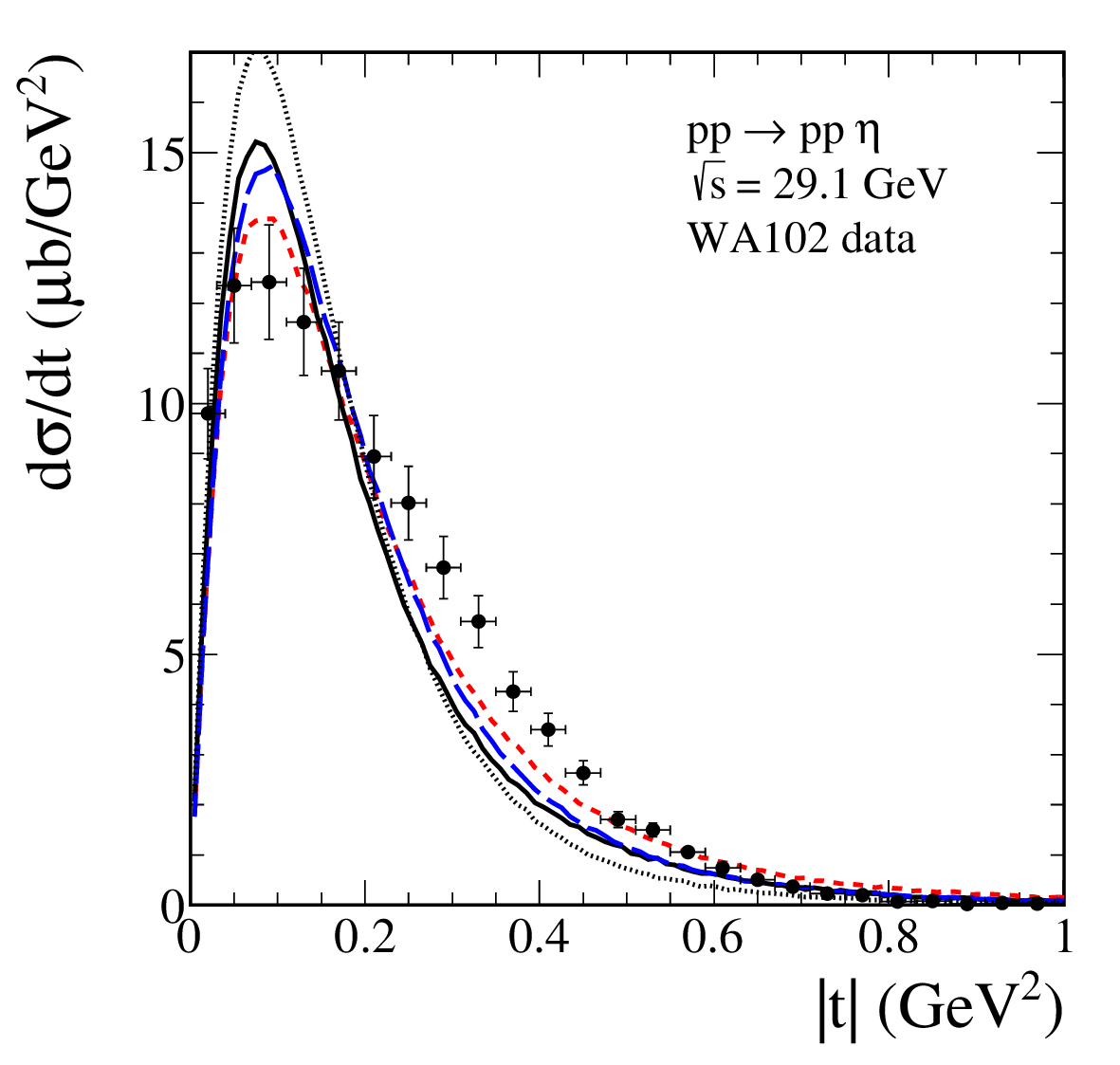}
\includegraphics[width=0.4\textwidth]{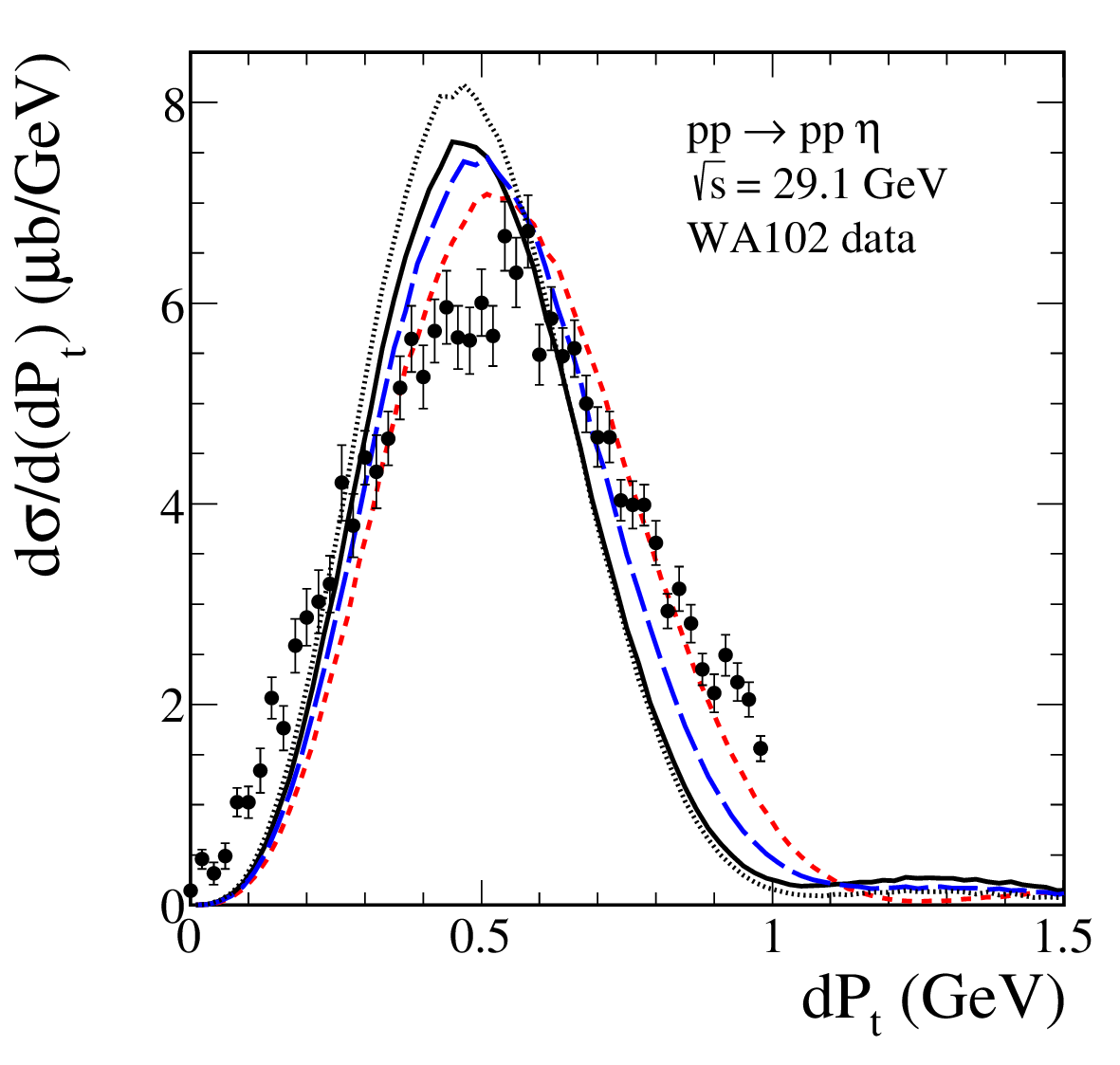}
\includegraphics[width=0.4\textwidth]{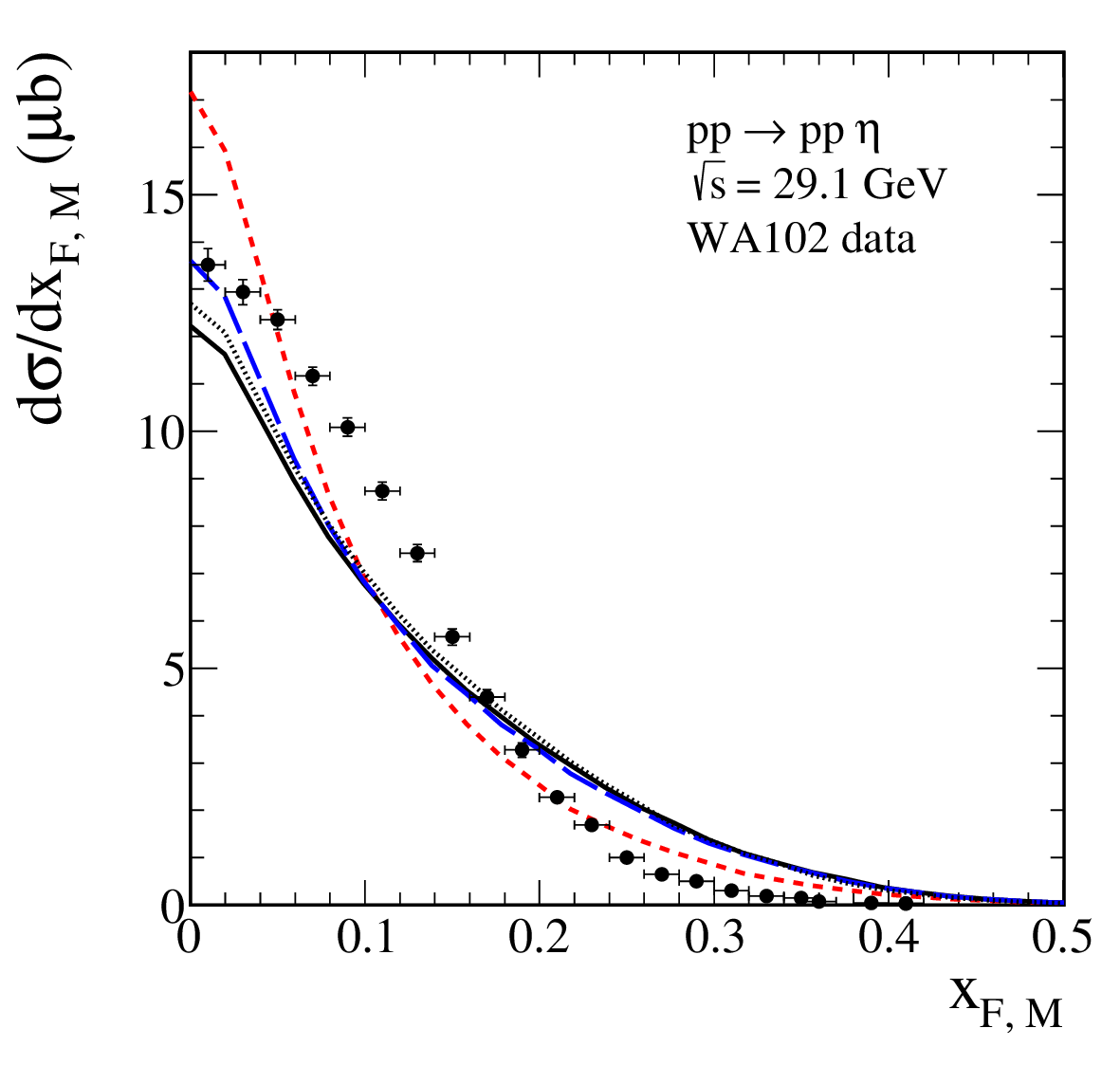}
  \caption{\label{fig:2}
  \small
Fits to the WA102 data \cite{WA102:1998ixr} 
for $\eta$ meson production at $\sqrt{s} = 29.1$~GeV.
The absorption effects are included in the calculations.
For the line description see Table~\ref{tab:parameters}.}
\end{figure}

In Fig.~\ref{fig:2} we show our fit results and the WA102 data
for the $pp \to pp \eta$ reaction at $\sqrt{s} = 29.1$~GeV.
The WA102 data points from \cite{WA102:1998ixr}
have been normalized to the mean value of the total cross section
(\ref{xs_eta}).
For the description of $\eta$ production 
the $\Pom \Pom$, $\Pom f_{2 \Reg}$, $f_{2 \Reg} \Pom$,
and $f_{2 \Reg} f_{2 \Reg}$
exchanges in the amplitude are included.
In the calculations we use four sets of parameters 
corresponding to Fits A--D
in Table~\ref{tab:parameters}.
In Set A, we assume that the $\Pom \Pom$-exchange contribution 
is more important than in others.
In Fit~C, $\Pom \Pom$ represents 40\%, 
$\Pom f_{2 \Reg}$ + $f_{2 \Reg} \Pom$ accounts for 32\%,
and $f_{2 \Reg} f_{2 \Reg}$ for 5\%.
In Fit~D we have 25\%, 48\%, and 4\%, respectively.

\begin{figure}[!ht]
\includegraphics[width=0.4\textwidth]{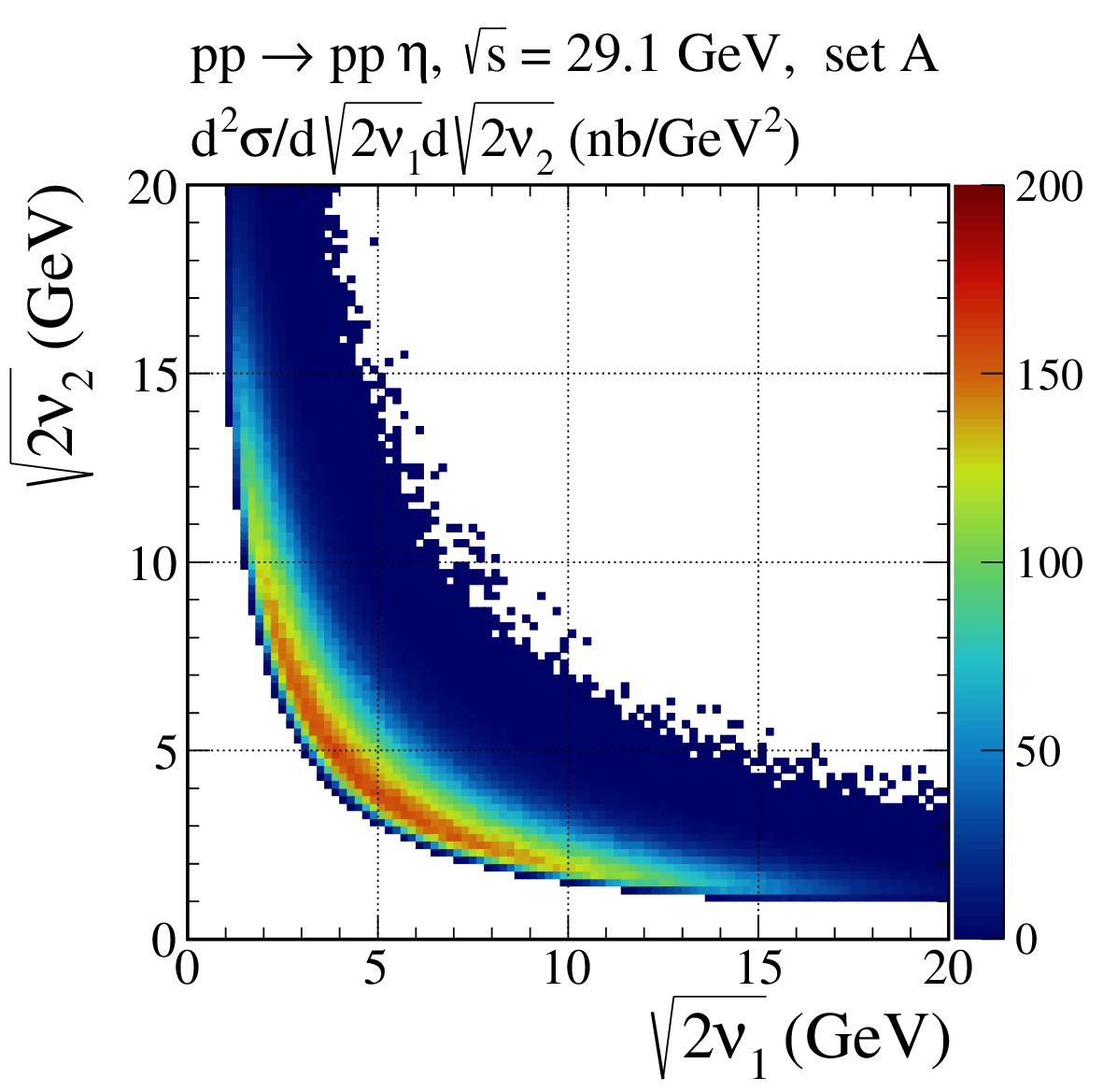}
\includegraphics[width=0.4\textwidth]{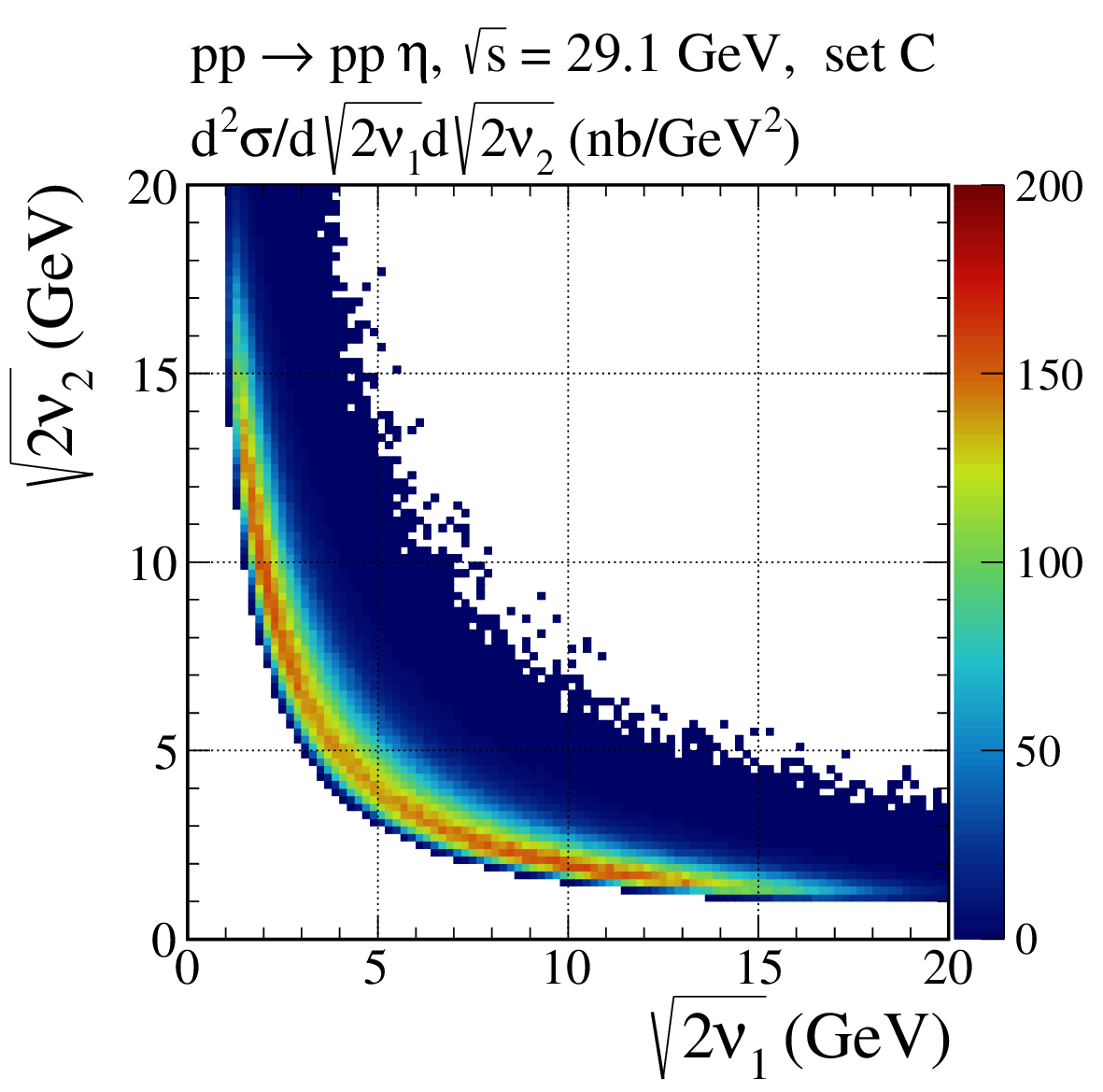}
\caption{\label{fig:2a}
\small
The two-dimensional distributions in ($\sqrt{2\nu_{1}}$, $\sqrt{2\nu_{2}}$)
for the $pp \to pp \eta$ reaction 
calculated for $\sqrt{s} = 29.1$~GeV.
The results for set~A (left panel) and set~C (right panel) are shown.}
\end{figure}

In Fig.~\ref{fig:2a} we show the two-dimensional distributions 
in ($\sqrt{2\nu_{1}}$, $\sqrt{2\nu_{2}}$).
The results for set~A (left panel) and set~C (right panel)
are presented.
We can see how these variables are correlated 
in the $\sqrt{2\nu_{1}}$-$\sqrt{2\nu_{2}}$ plane.
We find that the main contribution comes from the region where
$\sqrt{2\nu_{1}} \sqrt{2\nu_{2}} \approx 20$~GeV$^{2}$.
This is not far from the naive expectation
$\sqrt{2\nu_{1}} \sqrt{2\nu_{2}} \approx \sqrt{s_{1} s_{2}} \approx m_{\eta}\sqrt{s} \approx 16$~GeV$^2$;
see (D.17) of \cite{Lebiedowicz:2013ika}.

Now we summarize our analysis for the $pp \to pp \eta'$ 
and $pp \to pp \eta$ reactions
of this section. We have shown 
that double-pomeron exchange,
with suitable $\Pom \Pom \eta'$, $\Pom f_{2 \Reg} \eta' + f_{2 \Reg} \Pom  \eta'$,
and $f_{2 \Reg} f_{2 \Reg} \eta'$ couplings,
can give a reasonable description of the WA102 data
at $\sqrt{s} = 29.1$~GeV.
We have seen that in order to reproduce the shape of the $\phi_{pp}$ distribution with a maximum at $\phi_{pp} = \pi/2$, 
the two $(l,S) = (1,1)$ and $(3,3)$ couplings are needed 
with $g_{\Pom \Pom M}'' / g_{\Pom \Pom M}' \simeq 0.5$.
The dominant contribution to the cross section 
is from the coupling $(l,S) = (1,1)$.
If there are important contributions with
subleading $f_{2 \Reg}$-pomeron and $f_{2 \Reg}$-$f_{2 \Reg}$ exchanges,
the $\Pom \Pom \eta'$ couplings
could be significantly smaller.
In this case, our fits 5 and 6 suggest that 
$g_{\Pom \Pom M}'' / g_{\Pom \Pom M}' \simeq 1.0$.
For the $pp \to pp \eta$ reaction, we have found that the contributions of secondary exchanges are also significant.
In all cases
the interference terms between the different contributions
in the amplitudes
play an important role.

\subsection{Comment on the $\Pom \Pom \eta$ and $\Pom \Pom \eta'$ couplings obtained in the Sakai-Sugimoto model}
\label{sec:3aux}
In \cite{Anderson:2014jia,Anderson:2016zon} 
the following structure for the $\Pom \Pom M$ vertex 
(without a form-factor function and 
using our notation)
appears 
\begin{eqnarray}
i\Gamma_{\mu \nu,\kappa \lambda}^{\rm CS}(q_{1},q_{2}) =
2 i 
\left[
\left(\varkappa_{a} - \varkappa_{b} (q_{1} \cdot q_{2}) \right)
g_{\nu \lambda}
+ \varkappa_{b} \,q_{1 \lambda} q_{2 \nu}
\right]
\varepsilon_{\mu \kappa \rho \sigma}
q_{1}^{\rho} q_{2}^{\sigma}\,.
\label{vertex_CS}
\end{eqnarray}
The values of the coupling
constants $(\varkappa_{a}, \varkappa_{b}) = 
(0.084~{\rm GeV}^{-1}, 0.182~{\rm GeV}^{-3})$ were
determined using the Sakai-Sugimoto model 
in Sec.~IV of \cite{Anderson:2014jia}.
Note that in Table~I of \cite{Anderson:2016zon} 
smaller values for 
$(\varkappa_{a}, \varkappa_{b})
=$~(0.0222~GeV$^{-1}$, 0.0482~GeV$^{-3}$)
are quoted.
Nevertheless, the ratio is predicted to be 
$\varkappa_{b} / \varkappa_{a} \simeq 2.17$~GeV$^{-2}$.
The structure of the $\Pom \Pom M$ coupling is 
universal (model independent),
but the specific values of the coupling constants depend on
the details of the Sakai-Sugimoto model.
There is evidence that
the Sakai-Sugimoto model may underestimate
the values of the relevant coupling constants 
\cite{Anderson:2014jia,Anderson:2016zon}.

It is easy to see that the equivalence relation
\begin{eqnarray}
\Gamma^{\prime} + \Gamma'' = \Gamma^{\rm CS}
\label{equivalence1}
\end{eqnarray}
between the 'bare' vertices (\ref{vertex_pompomPS_11}),
(\ref{vertex_pompomPS_33}) and (\ref{vertex_CS})
holds if the respective coupling parameters satisfy 
the following conditions
\begin{eqnarray}
g'_{\Pom \Pom M}  = \varkappa_{a}\frac{M_{0}}{2}\,, 
\qquad
g''_{\Pom \Pom M} = \varkappa_{b}\frac{M_{0}^{3}}{4}\,.
\label{equivalence2}
\end{eqnarray}
In Sec.~\ref{sec:3a} we fix the 
coupling parameters
$g'_{\Pom \Pom M}$ and $g''_{\Pom \Pom M}$
from the comparison of the tensor-pomeron model 
with the WA102 experimental data.
From (\ref{equivalence2}) we have
\begin{eqnarray}
\frac{g''_{\Pom \Pom M}}{g'_{\Pom \Pom M}} 
=
\frac{M_{0}^{2}}{2} \frac{\varkappa_{b}}{\varkappa_{a}}\,,
\label{equivalence3}
\end{eqnarray}
and using the values of $\varkappa_{a}$ and $\varkappa_{b}$
from \cite{Anderson:2014jia,Anderson:2016zon}
we get
$g_{\Pom \Pom M}'' / g_{\Pom \Pom M}' \approx 1.1$.
This is in good agreement
with our findings from Fit 5 and Fit 6 
for the $\eta'$ production
as far as this ratio is concerned. 
But the magnitude of coupling constants
predicted in \cite{Anderson:2014jia,Anderson:2016zon} 
is much smaller than what we find in the tensor-pomeron approach
from the comparison with the WA102 data; see Table~\ref{tab:parameters}.

It may be that the WA102 energy is too low to draw
firm conclusions on the $\Pom \Pom \eta'$ couplings
due to very important contributions from reggeon-pomeron
and reggeon-reggeon exchanges. The study of CEP of $\eta'$
at LHC energies should answer this question since
there certainly $\Pom \Pom$ fusion is the dominant mechanism.

\subsection{Predictions for the LHC experiments}
\label{sec:3b}

First we present our results for
$pp \to pp \eta'$ for the LHC energy $\sqrt{s}=13$~TeV
where subleading reggeon contributions should be negligible,
at least, for the midrapidity region.
We consider parameter sets 1--6 
corresponding to fits 1--6 
as determined from the comparison with the WA102 data
(see Fig~\ref{fig:1} and Table~\ref{tab:parameters}).

\begin{figure}[!ht]
\includegraphics[width=0.4\textwidth]{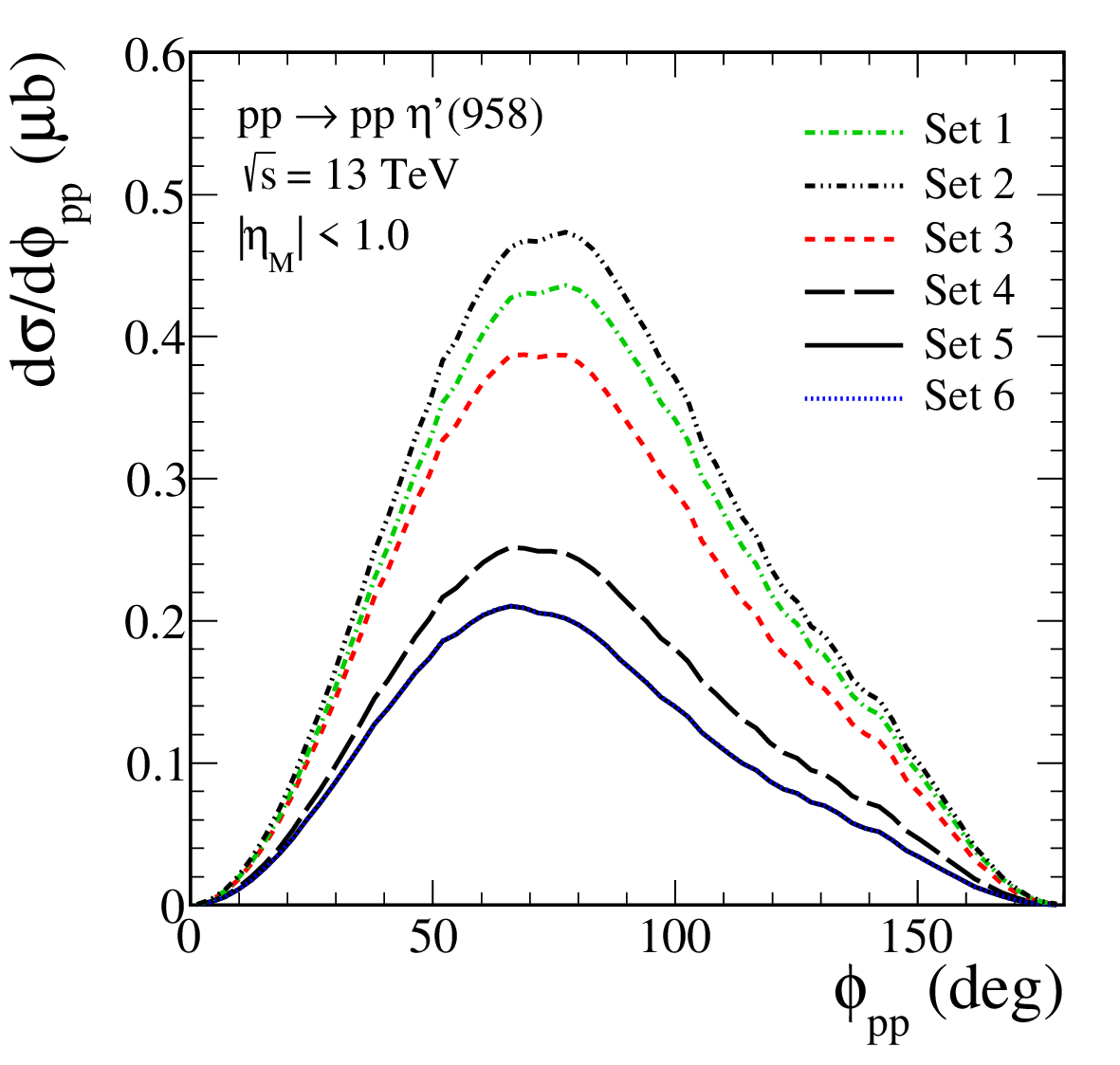}
\includegraphics[width=0.4\textwidth]{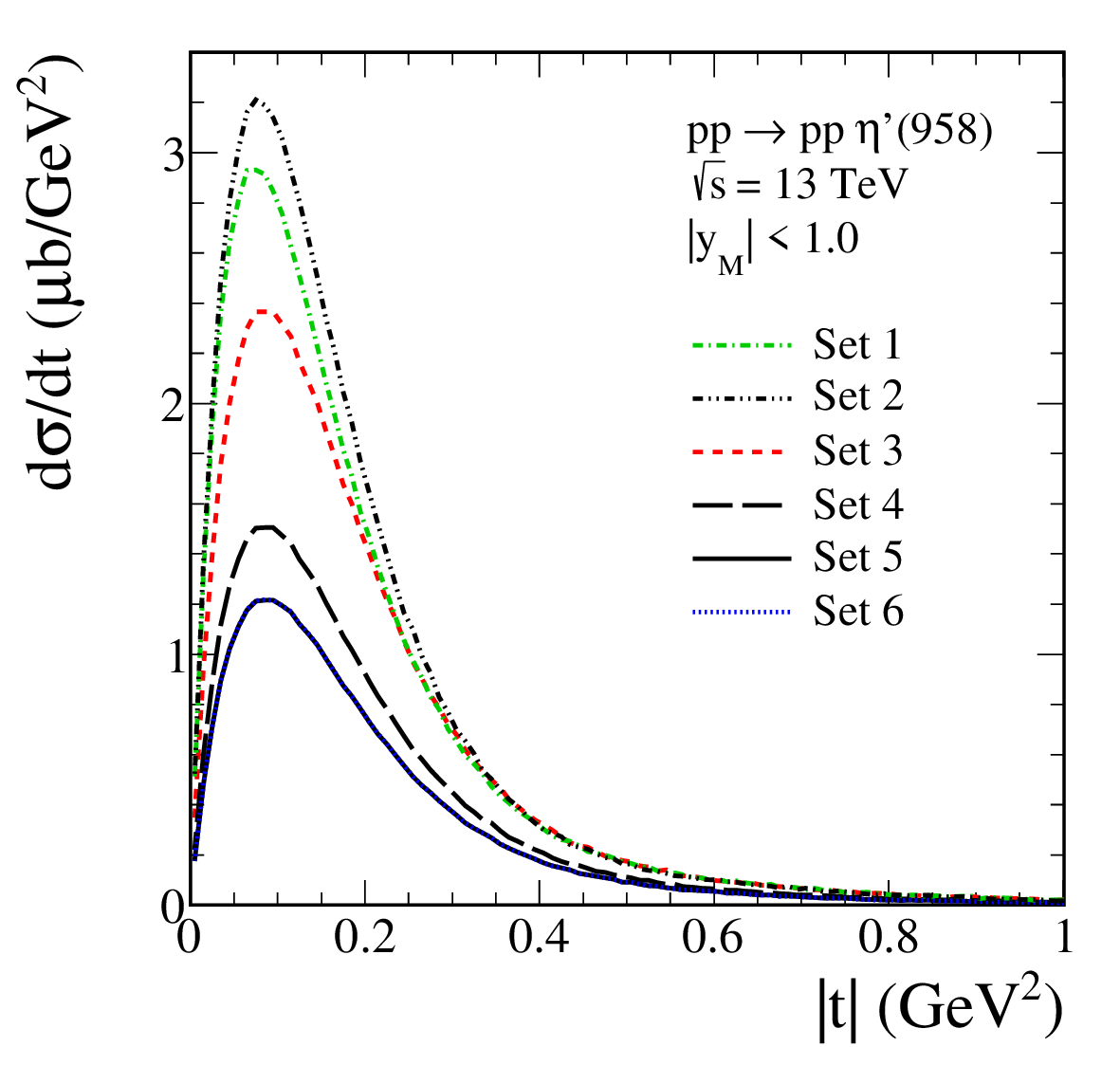}
\includegraphics[width=0.4\textwidth]{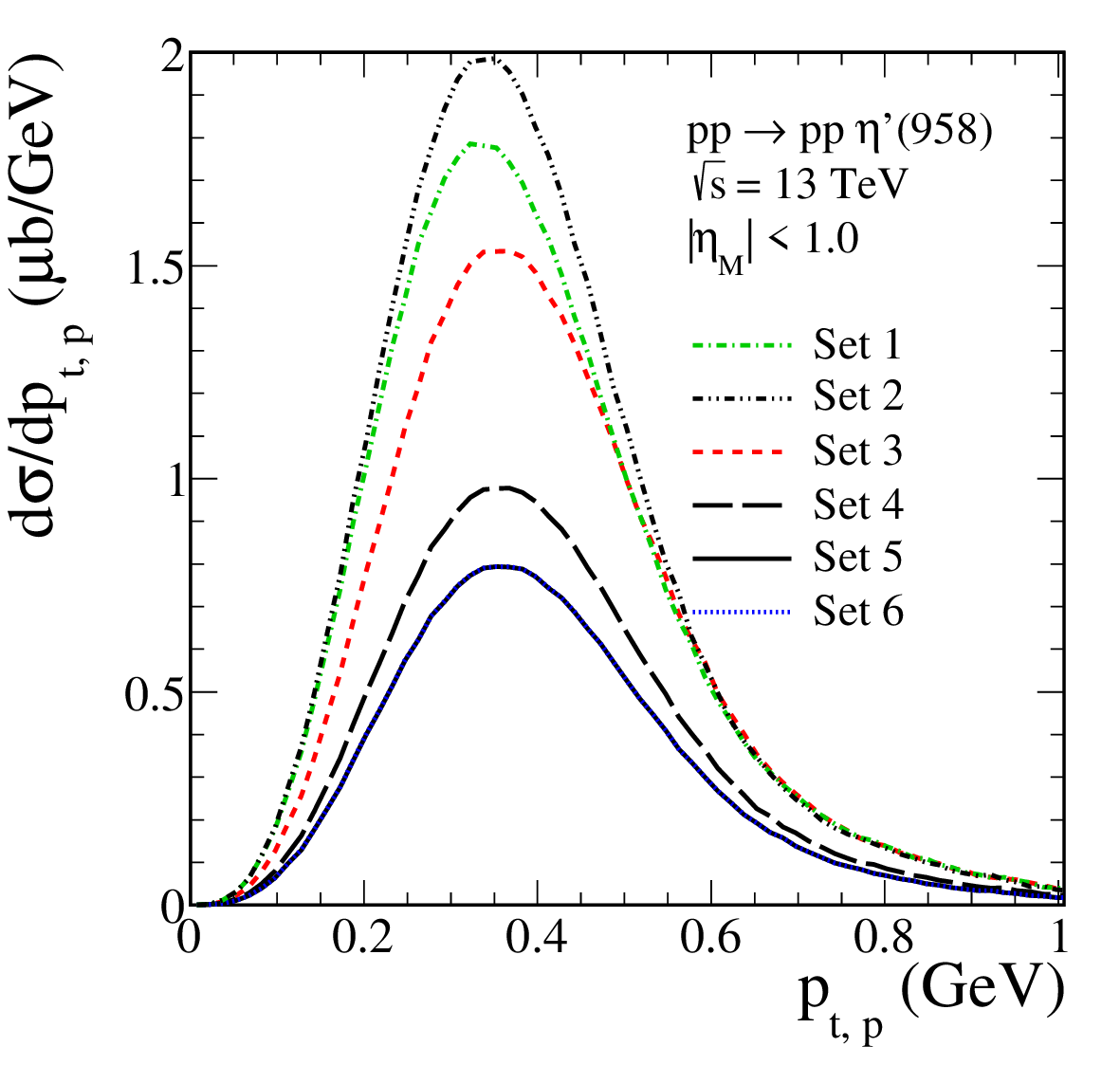}
\includegraphics[width=0.4\textwidth]{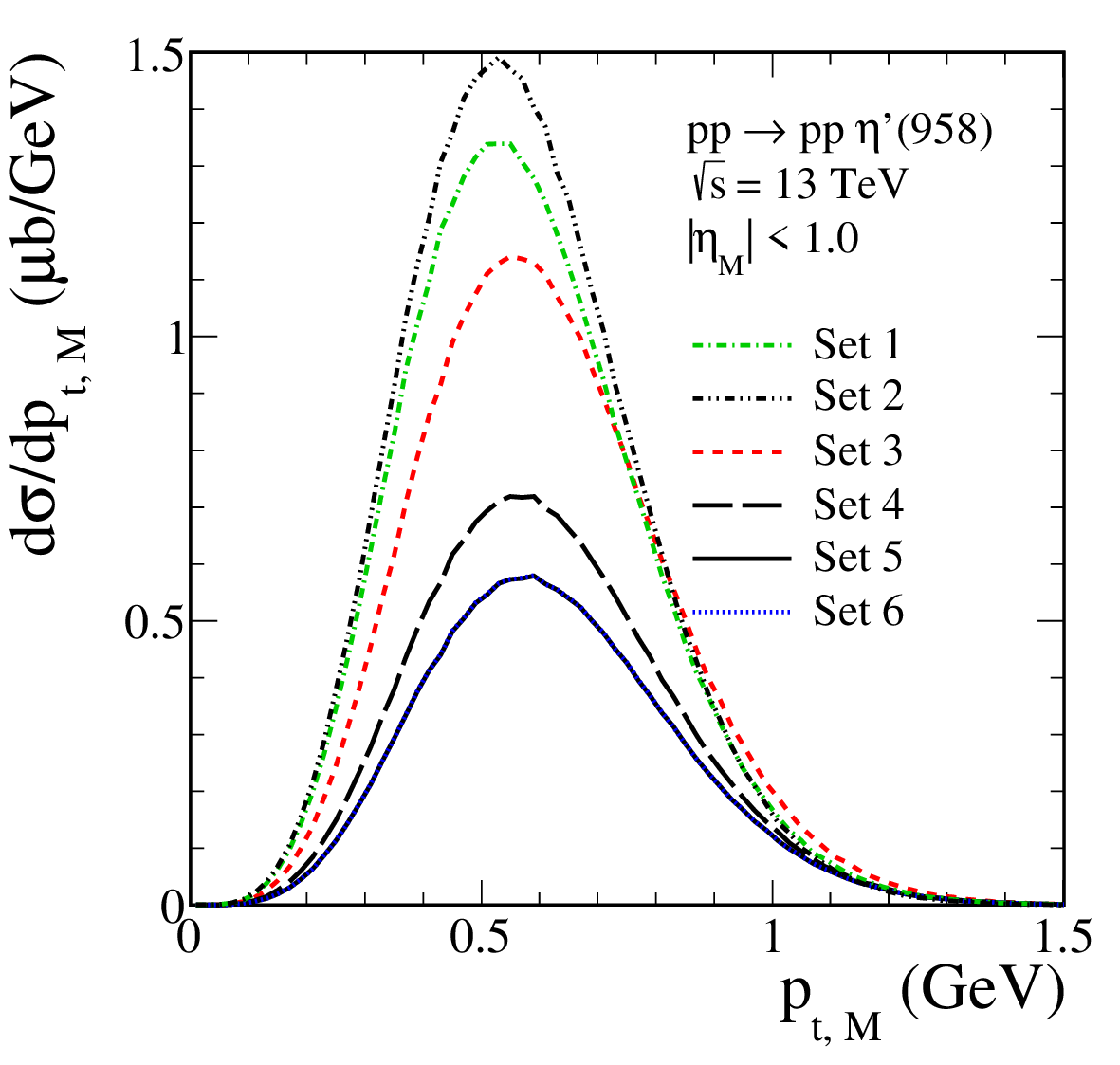}
  \caption{\label{fig:3}
  \small
The differential cross sections for the $pp \to pp \eta'$ reaction
calculated at $\sqrt{s} = 13$~TeV and for $|\eta_{M}| < 1.0$.
Here $\eta_{M}$ is the pseudorapidity of the $\eta'$ meson in the overall
c.m. system.
The results for the parameter sets 1--6 
(corresponding to fit 1--6 in Table~\ref{tab:parameters}) are shown.
The results for sets 5 and 6 
are practically identical
due to the small $f_{2 \Reg} f_{2 \Reg}$-exchange contribution at the LHC kinematics.
The absorption effects are included in the calculations.}
\end{figure}

Figure~\ref{fig:3} shows our results including a cut on 
the pseudorapidity of the $\eta'$ meson $|\eta_{M}| < 1.0$.
We see that $\eta'$ mesons are predominantly produced 
with the transverse momentum $p_{t, M} \simeq 0.5$~GeV 
and with momentum transfers between the protons of the order 
$p_{t, p} \sim \sqrt{|t_{1,2}|} \simeq 0.35$~GeV.
In all cases the absorption effects are included.
The main effect of absorption is a reduction of cross sections
by about 60\%
(see the values of $S^{2}$ in Table~\ref{tab:predictions_LHC} below).
It also modifies the differential distributions
because their shapes depend on the kinematics of the outgoing protons.
For instance, the maximum of the $\phi_{pp}$ distribution
is shifted from 90 degrees
towards smaller angles of about 70 degrees.

\begin{figure}[!ht]
\includegraphics[width=0.4\textwidth]{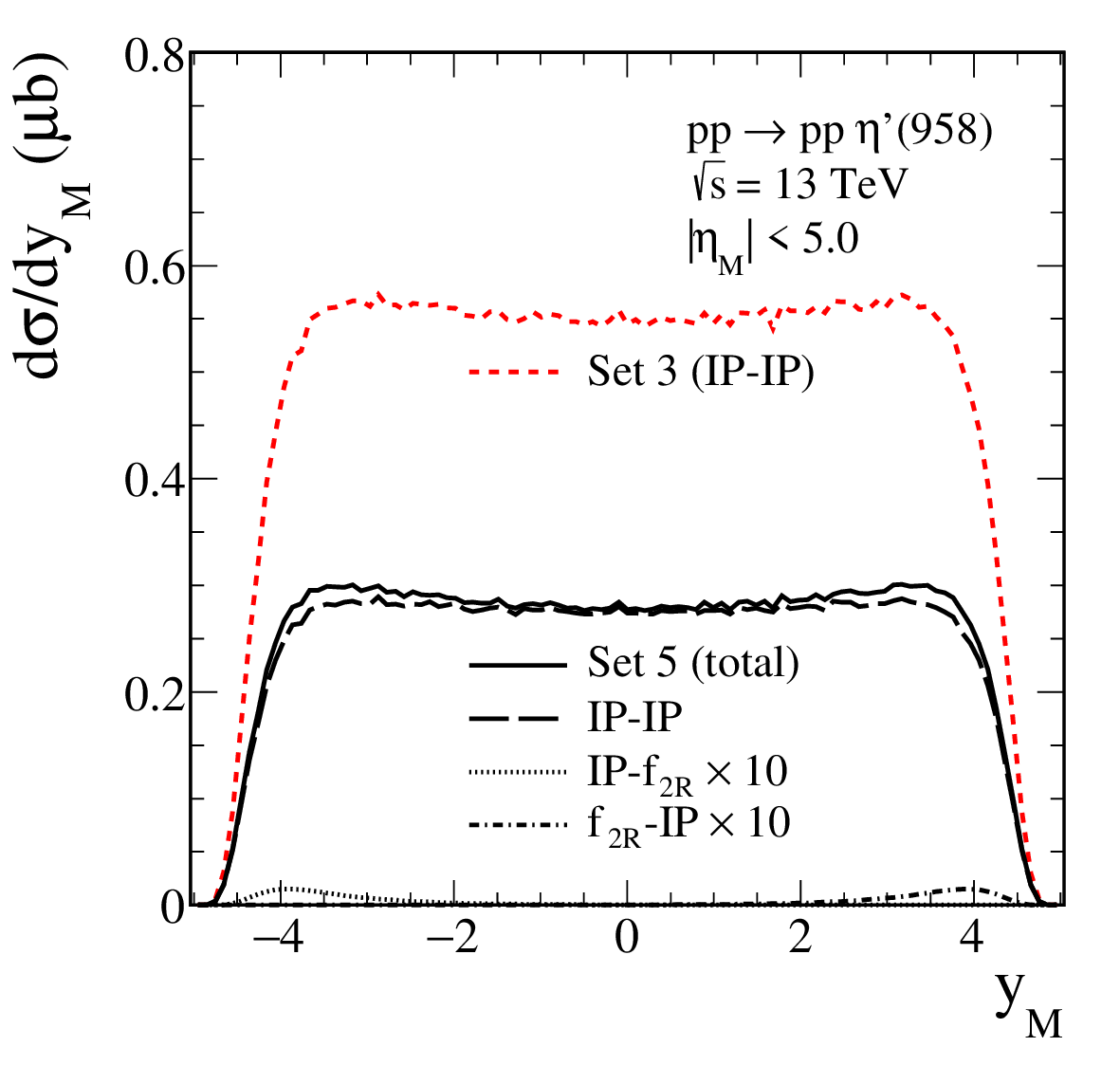}
\includegraphics[width=0.4\textwidth]{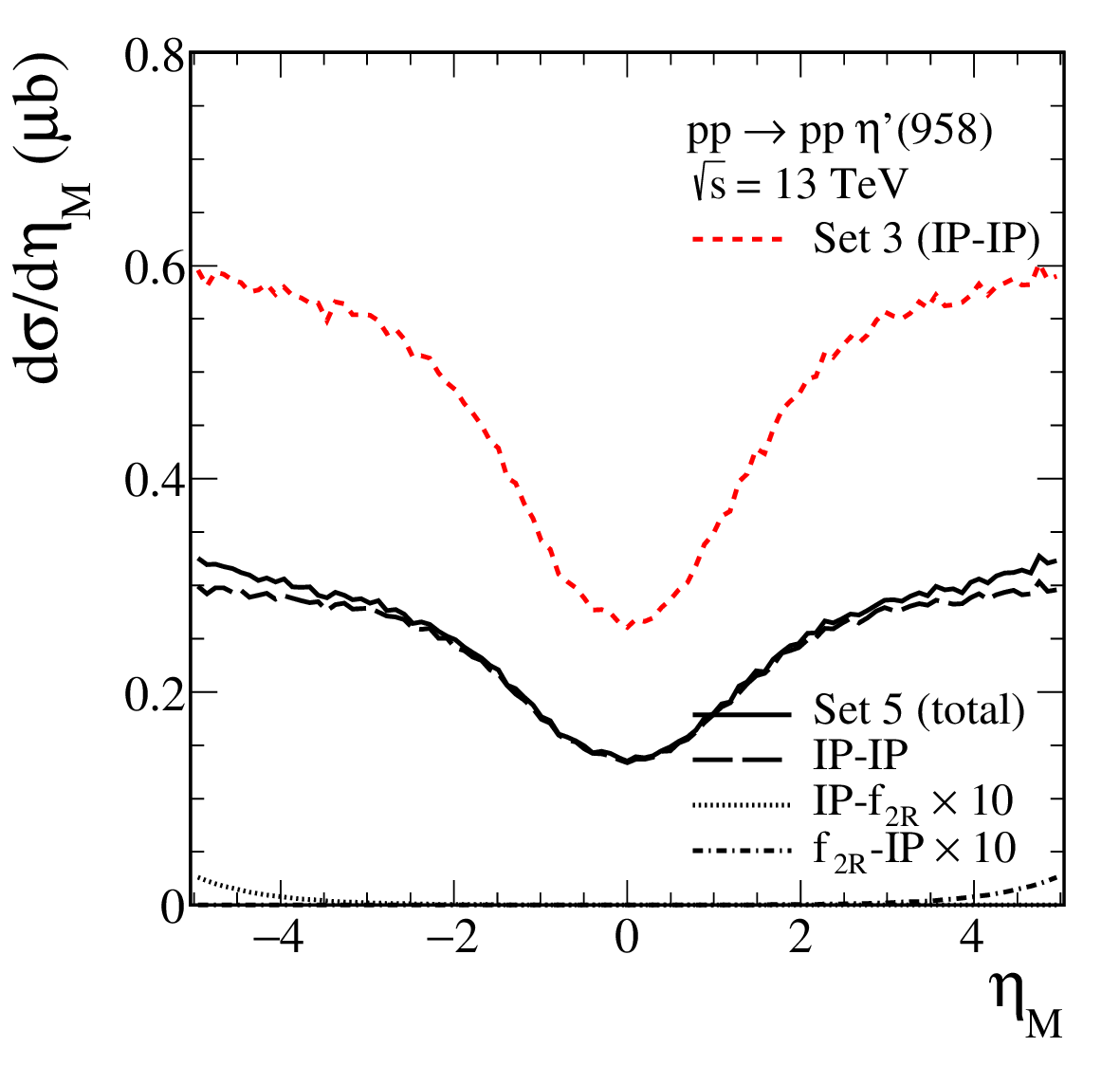}
  \caption{\label{fig:4}
  \small
The distributions in rapidity (${\rm y}_{M}$)
and in pseudorapidity ($\eta_{M}$) of the $\eta'$ meson
calculated for $\sqrt{s} = 13$~TeV.
The results for the parameter set 3 and 5 are shown.
The $\Pom f_{2 \Reg}$ and $f_{2 \Reg} \Pom$ contributions 
(in the case of set 5) 
are increased by a factor of 10 for visualization purposes.
The absorption effects are included in the calculations.}
\end{figure}

In Fig.~\ref{fig:4} we show the rapidity (${\rm y}_{M}$)
and pseudorapidity ($\eta_{M}$) distributions of the $\eta'$ meson.
The minimum in the pseudorapidity distribution 
at $\eta_{M} = 0$
can be understood as a kinematic effect; 
see Appendix~D of \cite{Lebiedowicz:2013ika}.
We can see that the pomeron-reggeon 
and reggeon-pomeron exchanges are separated and contribute
at backward and forward meson rapidity, respectively.
The interference of these terms with the dominant $\Pom \Pom$ exchange
produces enhancements of the cross section at ${\rm y}_{M} \gtrsim 3$.
This effect of secondary reggeon exchanges could be tested
by the LHCb Collaboration.

Figure~\ref{fig:4} clearly shows that
$\Pom f_{2 \Reg}$ fusion effects, which could be responsible
for a rather large fraction of the cross sections at the WA102 energy,
will be negligible at the LHC energy.
Thus, LHC experiments will settle the question of the magnitudes
of the $\Pom \Pom \eta'$ couplings.
The comparison of the fits to the WA102 
and the LHC data will be an important issue
since we will learn from it the magnitude of the subleading
exchange contributions.

\begin{figure}[!ht]
\includegraphics[width=0.4\textwidth]{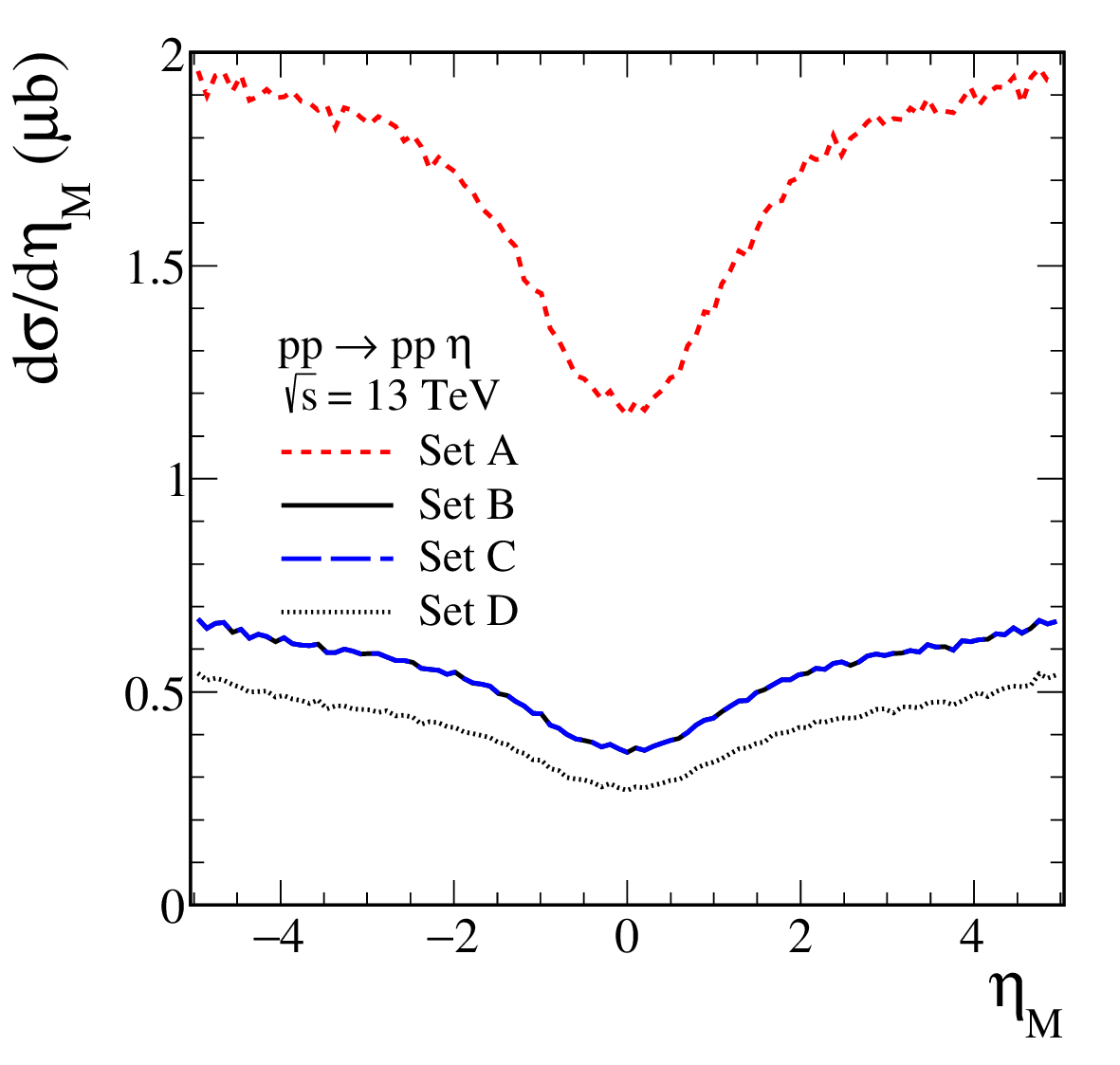}
\includegraphics[width=0.4\textwidth]{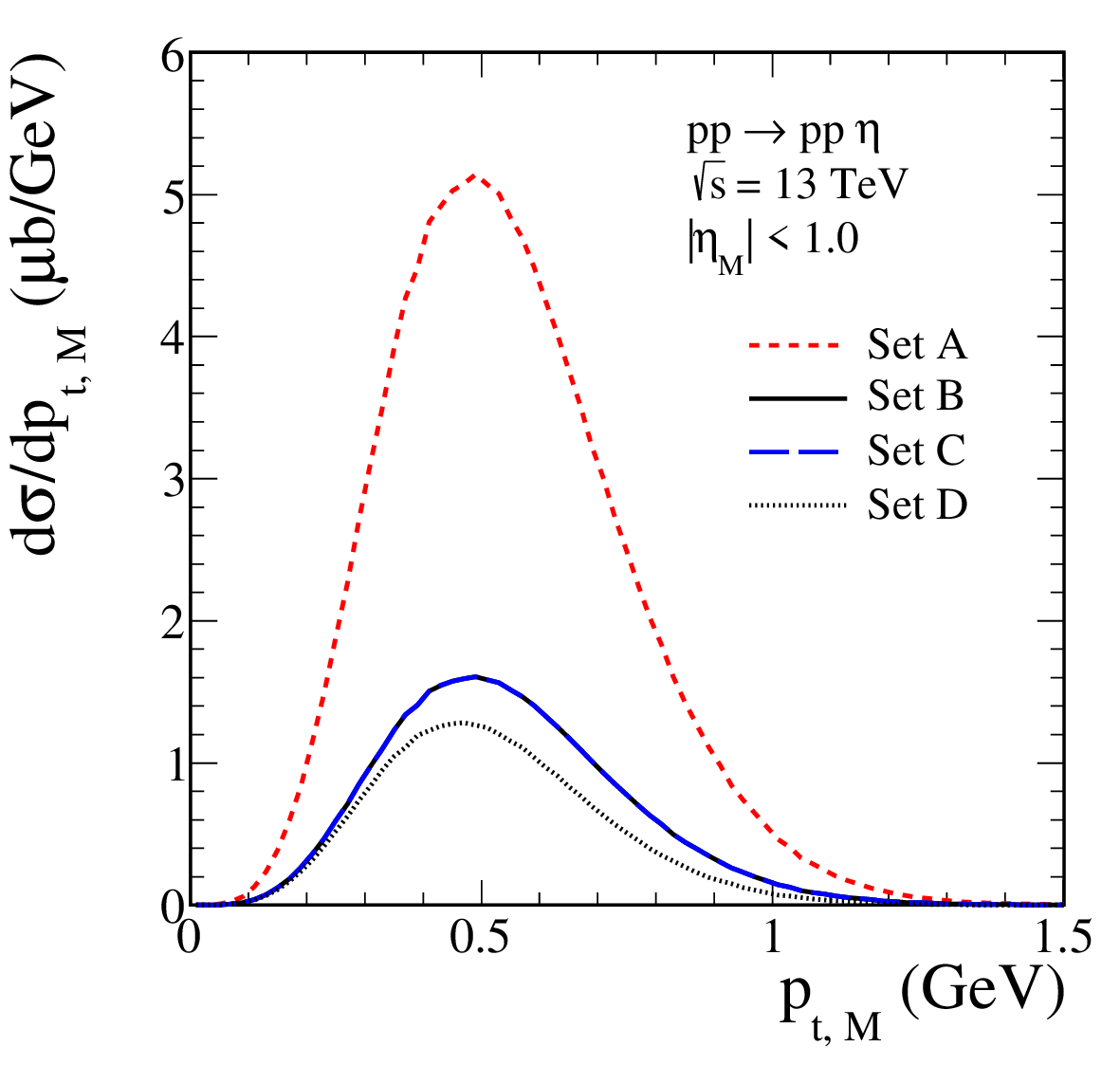}
\includegraphics[width=0.4\textwidth]{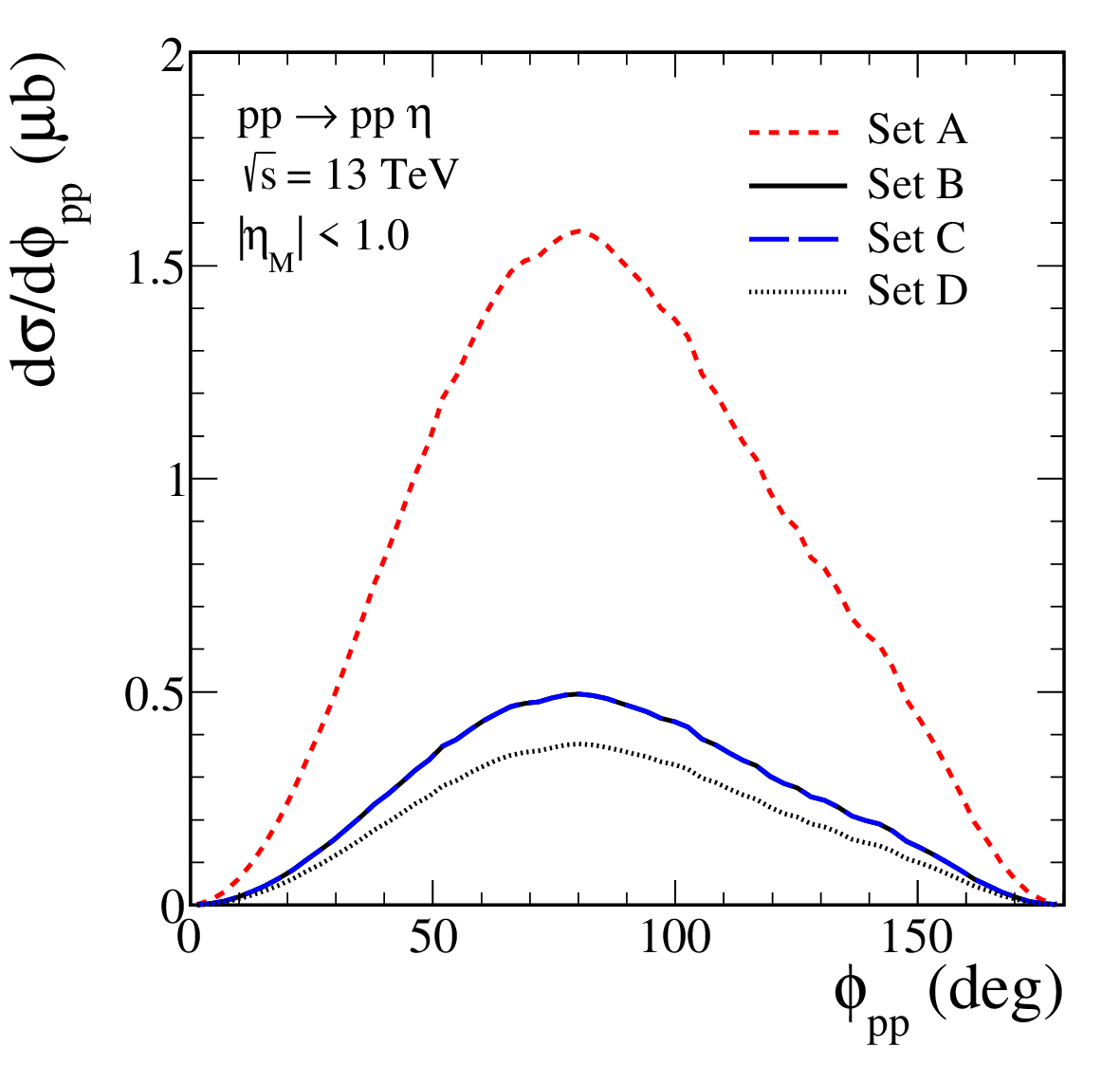}
\includegraphics[width=0.4\textwidth]{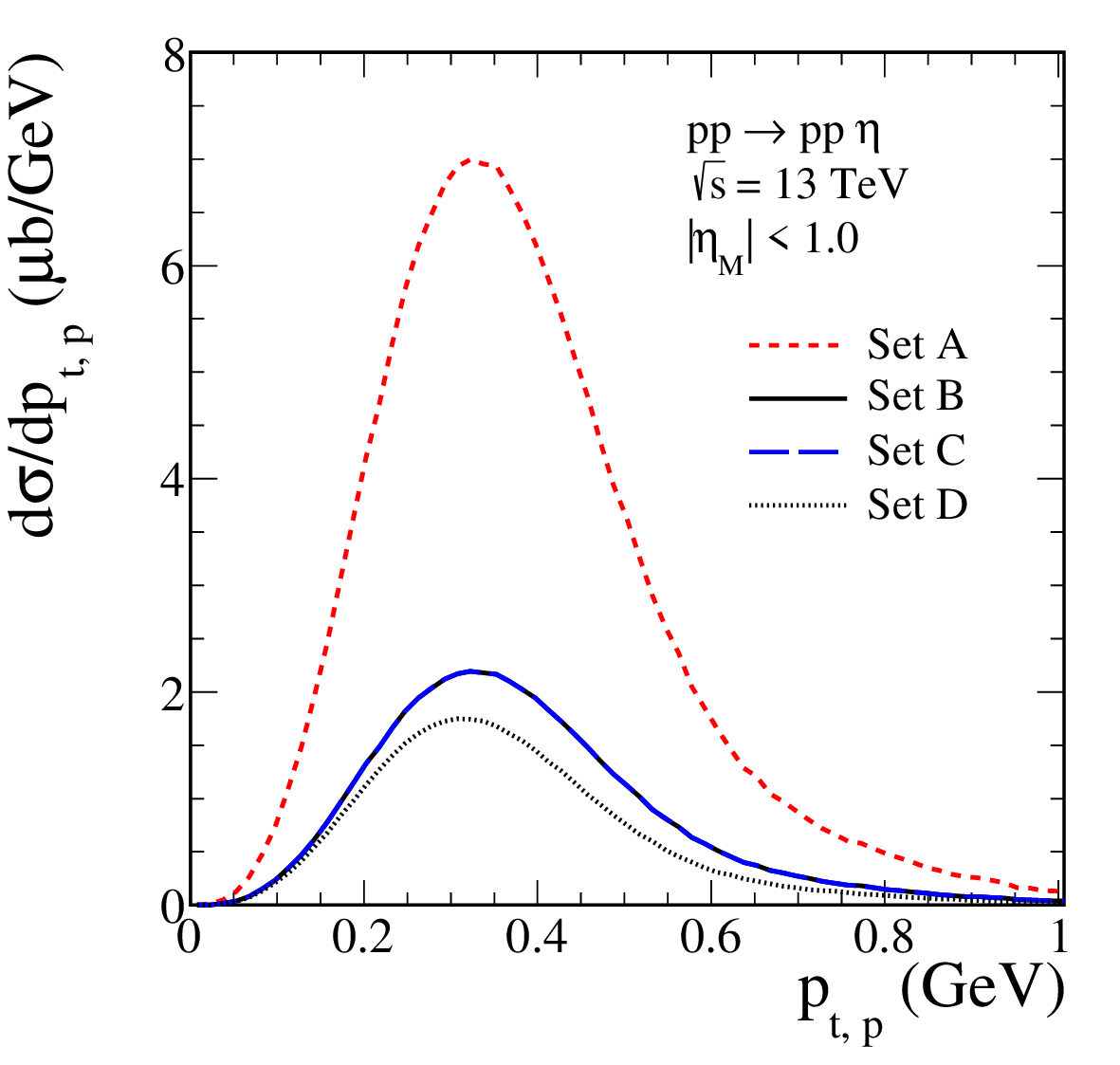}
  \caption{\label{fig:5}
  \small
The differential cross sections for the $pp \to pp \eta$ reaction
calculated at $\sqrt{s} = 13$~TeV 
with cut on $|\eta_{M}| < 1.0$.
The results correspond to the parameter sets A--D from Table~\ref{tab:parameters}.
The results for sets B and C 
are practically identical
due to the small $f_{2 \Reg} f_{2 \Reg}$-exchange contribution at the LHC kinematics.
The absorption effects are included in the calculations.}
\end{figure}

In Fig.~\ref{fig:5} we show our predictions
for the $pp \to pp \eta$ reaction
for the parameter sets A--D given in Table~\ref{tab:parameters}.
We see from Fig.~\ref{fig:2} and Table~\ref{tab:parameters}
that at the WA102 energy
we have fit~A to CEP of the $\eta$ meson
where the $\Pom \Pom \eta$ couplings are larger compared to fits~B--D.
Correspondingly, fit~A gives the largest cross sections for the LHC energy
in Fig.~\ref{fig:5}.
The~decision on the true magnitudes of the $\Pom \Pom \eta$ 
and the subleading couplings will again be given by the LHC experiments,
comparing the LHC data with the WA102 ones.

\begin{figure}[!ht]
\includegraphics[width=0.4\textwidth]{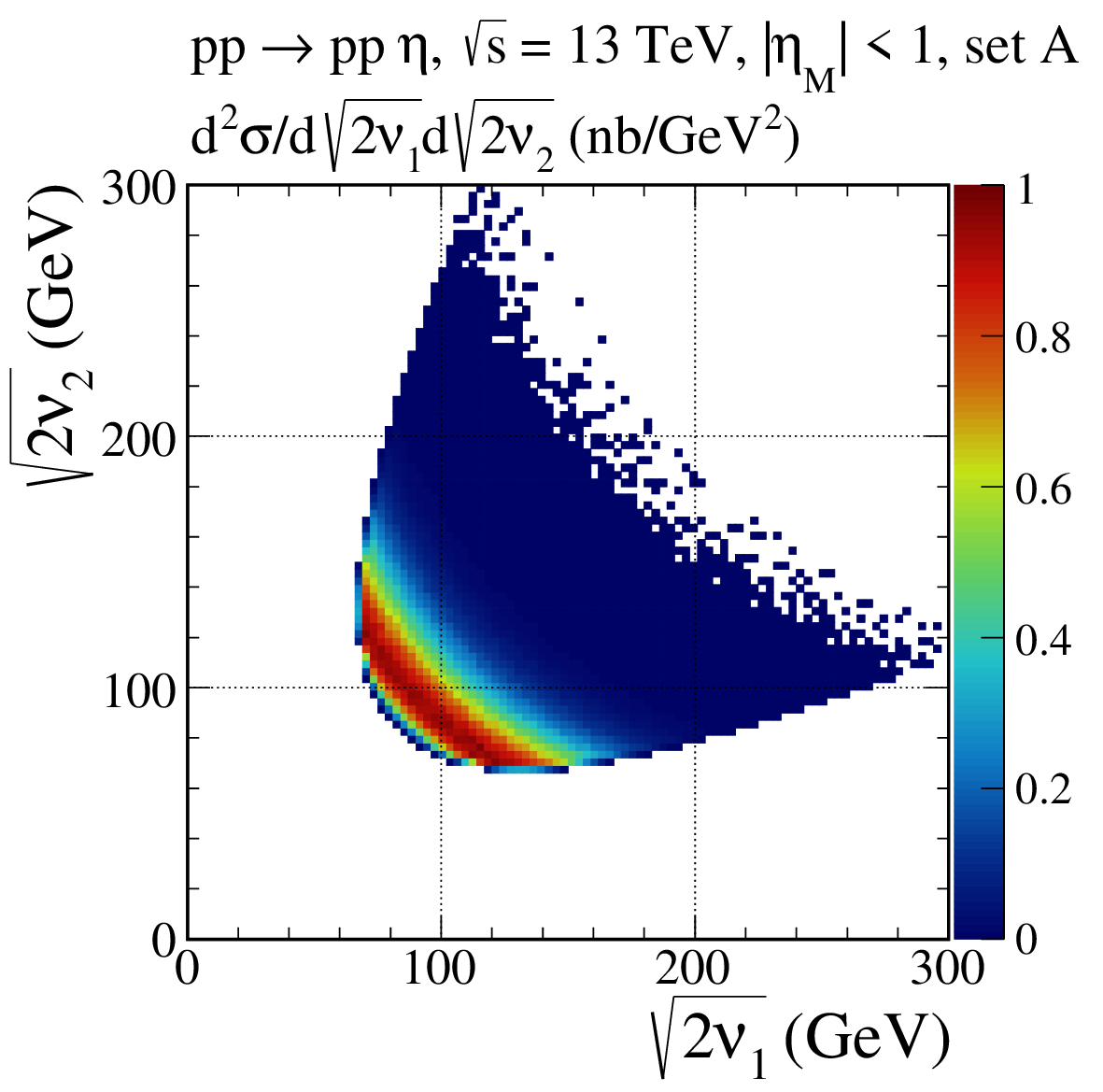}
\includegraphics[width=0.4\textwidth]{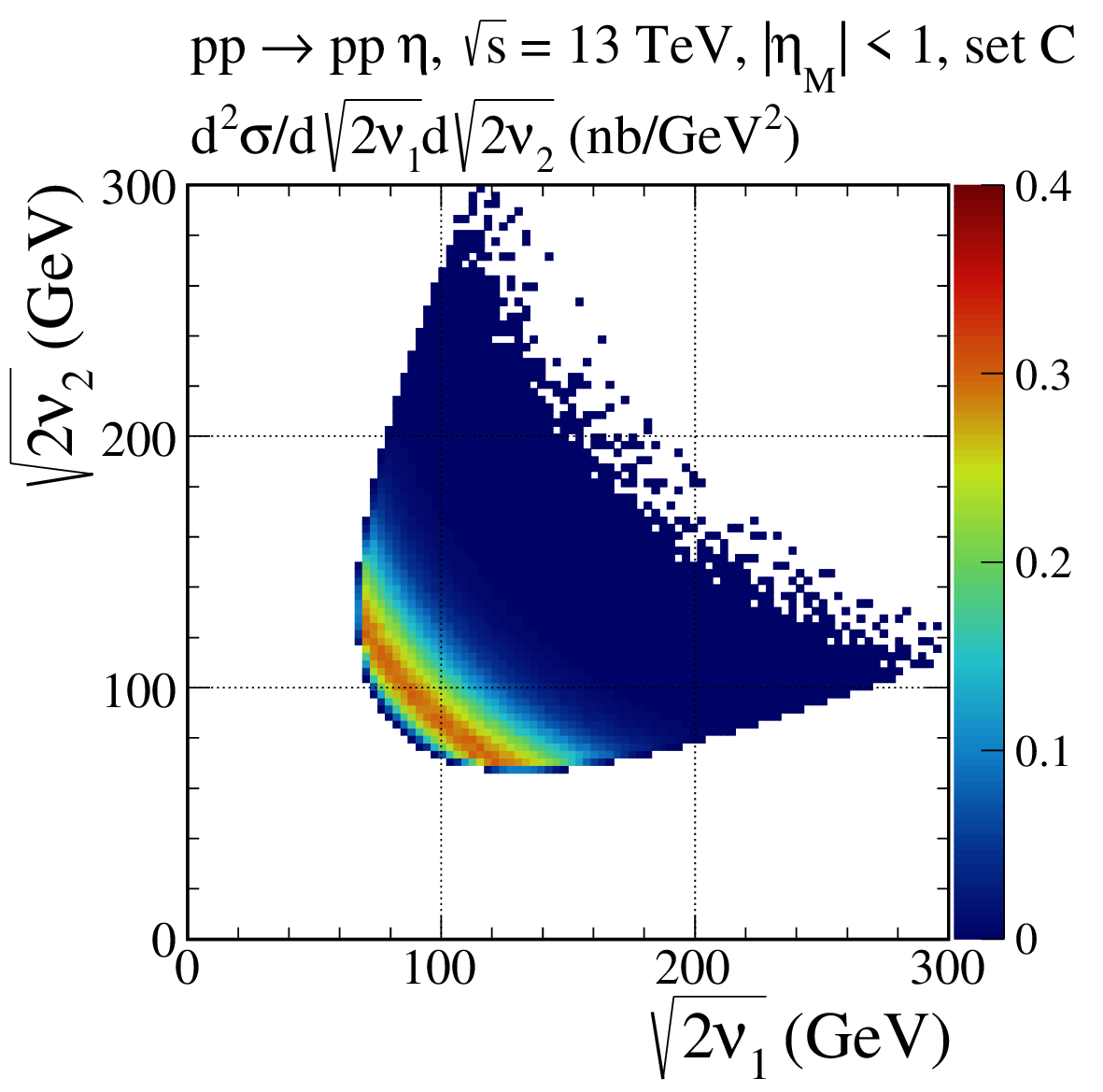}
\caption{\label{fig:6}
\small
The distributions in ($\sqrt{2\nu_{1}}$, $\sqrt{2\nu_{2}}$)
for the $pp \to pp \eta$ reaction 
calculated for $\sqrt{s} = 13$~TeV and with cut on $|\eta_{M}|<1$.
The~results for set~A (left panel) and set~C (right panel) are shown.}
\end{figure}

Figure~\ref{fig:6} shows the distributions
in ($\sqrt{2\nu_{1}}$, $\sqrt{2\nu_{2}}$) 
for the $pp \to pp \eta$ reaction.
Calculations were done for two parameter sets, A and C.
For the LHC kinematics, 
we have basically $\sqrt{2\nu_{1}} = \sqrt{s_{1}}$ and 
$\sqrt{2\nu_{2}} = \sqrt{s_{2}}$.
From $s_{1} s_{2} \approx m_{\eta}^{2} s$ we get
$\sqrt{2\nu_{1}} \sqrt{2\nu_{2}} \approx m_{\eta} \sqrt{s}
= 7.1 \times 10^{3}$~GeV$^{2}$.
On the other hand we find from Fig.~\ref{fig:6}
the main contribution to the cross section for
$\sqrt{2\nu_{1}} \sqrt{2\nu_{2}} \approx 9.0 \times 10^{3}$~GeV$^{2}$,
not far from the above value.
It is important to bear in mind that the energy variables 
$\nu_{1}$ and $\nu_{2}$
are correlated, as demonstrated in the figure.

In Table~\ref{tab:predictions_LHC} we have collected cross sections 
in $\mu$b for the reactions 
$pp \to pp \eta'$ and $pp \to pp \eta$ for $\sqrt{s}=13$~TeV
and for some kinematical cuts on the pseudorapidity of the mesons.
We must emphasize that our predictions 
for CEP of $\eta'$ and $\eta$
based on parameter sets 1--3 and set~A,
respectively,
(obtained assuming significant double-pomeron exchange 
at the WA102 energy $\sqrt{s} = 29.1$~GeV) 
should be considered as upper limits of the cross sections.
For sets~4--6, the cross section is smaller by a factor of about~2.
For the CEP of the $\eta$ meson, 
one can see a larger difference in the final results.
We can see from comparing the last columns 
of Tables~\ref{tab:parameters} and \ref{tab:predictions_LHC}
that the absorption effects 
are more important at the LHC than at the WA102 energy,
leading to a sizable reduction of the cross sections.
\begin{table}[!h]
\caption{
The integrated cross sections in $\mu$b for CEP of $\eta$
and $\eta'$ in $pp$ collisions for $\sqrt{s}=13$~TeV
for some kinematical cuts on the pseudorapidity of the mesons.
The results with absorption effects are presented.
In the calculations, 
we used the parameter sets corresponding to 
the fits 1--6 for $\eta'$ and to the fits A--D for $\eta$ 
from Table~\ref{tab:parameters}.}
\label{tab:predictions_LHC}
\begin{tabular}{c|c|c|c|c}
\hline
\hline
Meson $M$ 
& Cuts
& Parameter set 
& $\sigma_{\rm abs}$ ($\mu$b) 
& $S^{2}$\\
\hline
$\eta'(958)$ 
& $|\eta_{M}|<1.0$ 
 & 1 &  0.66 & 0.40\\
&& 2 &  0.72 & 0.42\\
&& 3 &  0.59 & 0.39\\
&& 4 &  0.37 & 0.40\\
&& 5 &  0.30 & 0.42\\
&& 6 &  0.31 & 0.42\\
\hline
& $2.0 < \eta_{M} < 5.0$ 
 & 1 &  1.94 & 0.40\\
&& 2 &  2.09 & 0.42\\
&& 3 &  1.67 & 0.39\\
&& 4 &  1.08 & 0.40\\
&& 5 &  0.88 & 0.42\\
&& 6 &  0.88 & 0.42\\
\hline
$\eta$ 
& $|\eta_{M}|<1.0$ 
 & A &  2.51 & 0.42\\
&& B &  0.78  & 0.42\\
&& C &  0.78  & 0.42\\
&& D &  0.59  & 0.46\\
\hline
& $2.0 < \eta_{M} < 5.0$ 
 & A & 5.58 & 0.42\\
&& B & 1.81 & 0.43\\
&& C & 1.81 & 0.43\\
&& D & 1.42 & 0.46\\
\hline
\hline
\end{tabular}
\end{table}

Let us finally comment on experimental signatures for our reactions
$pp \to pp \eta$ and $pp \to pp \eta'$.
One of the most prominent decay modes of the $\eta$ meson is 
$\pi^{+}\pi^{-}\pi^{0}$ with the branching fraction
${\cal BR}(\eta \to \pi^{+}\pi^{-}\pi^{0}) = (23.02 \pm 0.25) \%$ \cite{ParticleDataGroup:2024cfk}.
The final state $\pi^{+}\pi^{-}\pi^{0}$ is also
the main decay mode of the $\omega$ meson.
Recall that in the case of $pp \to pp \omega$ at LHC energy, 
the photoproduction (photon-pomeron fusion processes)
mechanism competes with the $\omega$-strahlung mechanism;
see e.g. \cite{Cisek:2011vt}.
From Fig.~10 of \cite{Cisek:2011vt} we can see 
that at midrapidities $|{\rm y}_{\omega}| \lesssim 2$
the photoproduction mechanism dominates over the $\omega$-strahlung one.
The cross section for photoproduction
$d\sigma(pp\to pp \omega) / d{\rm y}_{\omega}$ 
is estimated to be $\sim 0.1$~$\mu$b 
in the range $0 < {\rm y}_{\omega} < 5$.
Furthermore, experimental observation of 
a sharp increase of the cross section 
when going to large ${\rm y}_{\omega}$ would be a clear signature 
of the $\omega$-strahlung production.
In our opinion, the measurement of the relative contributions 
of the mesons $\eta$ and $\omega$
in the same system $\pi^{+} \pi^{-} \pi^{0}$
would provide very valuable information 
for comparing with model predictions.
Such results could be obtained by the ALICE and LHCb Collaborations
in experiments when the leading protons are not detected 
and instead only rapidity-gap conditions are checked experimentally.
The channel where the $\eta'(958)$ resonance
is to be observed is $\pi^{+}\pi^{-}\eta$,
and this channel is also prominent for the $f_{1}(1285)$ decay.
Here we predict much larger cross section
for CEP of $f_{1}(1285)$ via the $\Pom \Pom$-fusion process
(see Table~III of \cite{Lebiedowicz:2020yre})
than for the $\eta'$.
It should be kept in mind, however, 
that the theoretical results are sensitive to kinematics, 
i.e. depend on experimental cuts, 
as well as on the type of the $\Pom \Pom \eta'$ and $\Pom \Pom f_{1}$
couplings used in the calculation.
Clearly, all these topics deserve careful analyses, 
but they go beyond the scope of the present paper.

\section{Conclusions}
\label{sec:4}

In this paper we have discussed central exclusive production (CEP)
of $\eta$ and $\eta'$ mesons
in diffractive proton-proton collisions.
We have first shown that in a theory where the pomeron couples
like a scalar none of the particles $\eta$, $\eta'$, and $f_{1}(1285)$
can be produced by pomeron-pomeron fusion in CEP.
Therefore, observation of CEP of any of these particles
at the LHC will be a striking evidence against a scalar character
of the pomeron.

We have then discussed in detail,
using the tensor-pomeron model,
$\eta$ and $\eta'$ CEP at the WA102 energy $\sqrt{s} = 29.1$~GeV
and at the LHC energy $\sqrt{s} = 13$~TeV.
We considered the $\Pom \Pom$, the $\Pom f_{2 \Reg} + f_{2 \Reg} \Pom$, 
and the $f_{2 \Reg} f_{2 \Reg}$ fusion processes
giving $\eta$ and $\eta'$.
At the low WA102 energy our fits to the data
did not give us a clear indication of the relative size of
the $\Pom \Pom$, $\Pom f_{2 \Reg} + f_{2 \Reg} \Pom$, 
and $f_{2 \Reg} f_{2 \Reg}$ contributions.
We have shown that in contrast to this experiments
at the LHC should be able to give us precise numbers
for the $\Pom \Pom \eta$ and $\Pom \Pom \eta'$ coupling parameters.

Finally we discuss the coupling of two pomerons to the $\eta$ meson
from the point of view of SU(3)$_{\rm F}$-flavor symmetry.
If SU(3)$_{\rm F}$ was an exact symmetry of QCD the $\eta$ meson
would be a pure SU(3)$_{\rm F}$ octet state,
the pomeron a pure SU(3)$_{\rm F}$ singlet gluonic object
and consequently the coupling $\Pom \Pom \eta$ would be zero.
But SU(3)$_{\rm F}$ is badly broken.
There is octet-singlet mixing and the physical $\eta$
contains both octet and singlet components;
see Chapter~15.3 of \cite{ParticleDataGroup:2024cfk} for a review.
Also, there is the possibility of a purely gluonic part 
in the wave function of the $\eta$;
see for instance \cite{Diehl:2001dg,Kroll:2002nt,Kroll:2021zss}.
A direct observation of SU(3)$_{\rm F}$ symmetry breaking
in the pomeron-meson-meson coupling is provided
by the comparison of $\pi p$ and $K p$ total cross sections
which are, for instance, 
shown in Figs.~3.1 and 3.2 of \cite{Donnachie:2002en}.
From a fit to these cross sections the pomeron parts are found to be
\begin{eqnarray}
\sigma(\pi p)|_{\Pom} &=& 13.63 \, (s M_{0}^{-2})^{\epsilon_{\Pom}}\,, \nonumber \\
\sigma(K p)|_{\Pom} &=& 11.93 \, (s M_{0}^{-2})^{\epsilon_{\Pom}}\,, \nonumber \\
M_{0} &=& 1~{\rm GeV}\,, \quad \epsilon_{\Pom} = 0.0808\,.
\label{4.1}
\end{eqnarray}
This gives direct evidence for SU(3)$_{\rm F}$ breaking
in the pomeron-meson-meson coupling.
But this can be understood quantitatively in the calculations
based on the functional-integral approach to soft high-energy 
hadron-hadron scattering introduced in \cite{Nachtmann:1991ua};
see \cite{Nachtmann:Lectures} and Chapter~8 of \cite{Donnachie:2002en} for reviews.
There, hadron-hadron scattering is calculated from
the correlation function of lightlike Wegner-Wilson loops representing the hadrons.
The results depend, of course, on the transverse size of the loops.
This model gives a dependence of the pomeron part
of the total cross sections on the hadron radii
$R_{h}$ divided by $R_{p}$.
See Fig.~8.11 of \cite{Donnachie:2002en}
where this is shown for
$\sigma(h p)|_{\Pom}$
with $h = p, \pi, K$, and $J/\psi$.
In this approach, the size of the strings between the quark and the antiquark in the meson
determines the coupling to the pomeron.
The quark and antiquark at the endpoints of the string
determine the SU(3)$_{\rm F}$ assignment
of the meson but not directly its coupling to the pomeron. We are looking forward to the measurements
of CEP of $\eta$ and $\eta'$ mesons at the LHC.
These will tell us if in the $\Pom \Pom \eta$ and $\Pom \Pom \eta'$ couplings the string extensions
or the flavor quantum numbers at the string
endpoints of the mesons are the relevant parameters.
A comparatively large $\Pom \Pom \eta$ coupling,
as we advocate in our present paper,
would be support for the former possibility.

\acknowledgments
The authors would like to thank C. Ewerz, R. McNulty,
and R. Schicker for useful discussions.
The work of A.S. was partially supported by the Centre for Innovation and
Transfer of Natural Sciences and Engineering Knowledge in Rzesz\'ow (Poland).

\appendix

\section{The pomeron-$f_{2 \Reg}$-pseudoscalar-meson coupling}
\label{sec:A1}

Here we discuss the couplings $\Pom f_{2 \Reg} M$ 
where $M$ denotes a pseudoscalar meson.
We proceed as in Appendix~A of \cite{Lebiedowicz:2013ika}.
We consider first the fictitious reaction 
of a ``real pomeron'' of spin~2
and a ``real $f_{2 \Reg}$ reggeon'' of spin~2 
giving the meson $M$.
Using the notation as in (A.10) of \cite{Lebiedowicz:2013ika}
we have in the rest system of $M$
\begin{eqnarray}
&& \Pom(\bk,2,m_{1}) + f_{2 \Reg} (-\bk,2,m_{2}) \to M(J,J_{z}) \,, \nonumber\\
&& J=0\,, 
\quad J_{z}=0 \,, 
\quad m_{1,2} \in \lbrace -2, \ldots, 2 \rbrace \,.
\label{A1}
\end{eqnarray}
Here $\bk$ is the momentum,
$J$ the total angular momentum of $M$,
and
$m_{1}$, $m_{2}$, and $J_{z}$ the 
$z$-components of the angular momenta.
We combine the spins of $\Pom$ and $f_{2 \Reg}$ to the total spin
\begin{eqnarray}
S = 0, 1, 2, 3, 4 \,.
\label{A2}
\end{eqnarray}
Then we combine $S$ with the angular momentum $l = 0, 1, 2, \ldots$
to give us $J$,
the total spin of the system.
The parity of the state, obtained in complete analogy to
(A.11)--(A.13) and (A.15), (A.16), of \cite{Lebiedowicz:2013ika}
\begin{eqnarray}
\ket{l, S; J, J_{z}}
\label{A3}
\end{eqnarray}
is then $(-1)^{l}$.
Clearly, for $S$ from (\ref{A2}) we can get $J = 0$
only for the following combinations $(l,S)$
shown in Table~\ref{tab:table3}.
\begin{table}[!h]
\caption{Values of orbital angular momentum $l$ and total spin $S$
which can lead in (\ref{A1}) to a state with total spin
$J=0$. $P$ is the parity of this state.}
\label{tab:table3}
\begin{tabular}{c|c|c|c}
\hline
\hline
$l$ & $S$ & $J$ & $P$ \\
\hline
$0$ & $0$ & $0$ & $+$ \\
$\textbf{1}$ & $\textbf{1}$ & $0$ & $-$ \\
$2$ & $2$ & $0$ & $+$ \\
$\textbf{3}$ & $\textbf{3}$ & $0$ & $-$ \\
$4$ & $4$ & $0$ & $+$ \\
\hline
\hline
\end{tabular}
\end{table}
We get only the combinations $(l,S) = (1,1)$ and $(3,3)$
which give us a pseudoscalar state.

Now we construct the coupling Lagrangians corresponding
to these couplings.
For $(l,S) = (1,1)$ we set as in (2.3) of \cite{Lebiedowicz:2013ika}
with $\tilde{\chi}_{M}(x)$ the pseudoscalar meson field
\begin{eqnarray}
{\cal L}_{\Pom f_{2 \Reg} M}'(x) &=& 
-\dfrac{4}{M_{0}}  g_{\Pom f_{2 \Reg} M}'  
\left[ \partial_{\rho} \Pom_{\mu \nu}(x) \right] 
\left[ \partial_{\sigma} f_{2 \Reg\,\kappa \lambda}(x) \right] 
g^{\mu \kappa} \varepsilon^{\nu \lambda \rho \sigma}
\tilde{\chi}_{M}(x) 
\nonumber \\
&=&
-\dfrac{2}{M_{0}} g_{\Pom f_{2 \Reg} M}'  
\bigg{\lbrace}
\left[ \partial_{\rho} \Pom_{\mu \nu}(x) \right] 
\left[ \partial_{\sigma} f_{2 \Reg\,\kappa \lambda}(x) \right] 
g^{\mu \kappa} \varepsilon^{\nu \lambda \rho \sigma} 
+
\left[ \partial_{\rho} f_{2 \Reg\, \mu \nu}(x) \right] 
\left[ \partial_{\sigma} \Pom_{\kappa \lambda}(x) \right] 
g^{\mu \kappa} \varepsilon^{\nu \lambda \rho \sigma}
\bigg{\rbrace}
\tilde{\chi}_{M}(x) \,.  \qquad
\label{A4}
\end{eqnarray}
Here $\Pom_{\mu \nu}(x)$ and $f_{2 \Reg\,\mu \nu}(x)$
are the effective pomeron and $f_{2 \Reg}$-reggeon field operators, respectively.\newline
For the $(l,S) = (3,3)$ coupling Lagrangian we set as in
(2.5) of \cite{Lebiedowicz:2013ika}
\begin{eqnarray}
{\cal L}_{\Pom f_{2 \Reg} M}''(x) &=& 
-\dfrac{2 g_{\Pom f_{2 \Reg} M}''}{M_{0}^{3}} 
\varepsilon^{\mu_{1} \mu_{2} \nu_{1} \nu_{2}} 
\left( \partial_{\mu_{1}} \tilde{\chi}_{M}(x) \right)
\left[ 
\big{(} 
\partial_{\mu_{3}} \Pom_{\mu_{4} \nu_{1}}(x) -
\partial_{\mu_{4}} \Pom_{\mu_{3} \nu_{1}}(x) 
\big{)} 
\twosidep{\mu_{2}} 
\big{(} 
\partial^{\mu_{3}} f_{2 \Reg \quad \nu_{2}}^{\quad \mu_{4}}(x) -
\partial^{\mu_{4}} f_{2 \Reg \quad \nu_{2}}^{\quad \mu_{3}}(x)
\big{)}  
\right] 
\nonumber \\
&=&
-\dfrac{g_{\Pom f_{2 \Reg} M}''}{M_{0}^{3}} 
\varepsilon^{\mu_{1} \mu_{2} \nu_{1} \nu_{2}} 
\left( \partial_{\mu_{1}} \tilde{\chi}_{M}(x) \right)
\Big{[}
\big{(} 
\partial_{\mu_{3}} \Pom_{\mu_{4} \nu_{1}}(x) -
\partial_{\mu_{4}} \Pom_{\mu_{3} \nu_{1}}(x) 
\big{)} 
\twosidep{\mu_{2}} 
\big{(} 
\partial^{\mu_{3}} f_{2 \Reg \quad \nu_{2}}^{\quad \mu_{4}}(x) -
\partial^{\mu_{4}} f_{2 \Reg \quad \nu_{2}}^{\quad \mu_{3}}(x) \big{)} 
\nonumber \\
&& + \,
\big{(} 
\partial_{\mu_{3}} f_{2 \Reg \, \mu_{4} \nu_{1}}(x) -
\partial_{\mu_{4}} f_{2 \Reg \, \mu_{3} \nu_{1}}(x) \big{)} 
\twosidep{\mu_{2}} 
\big{(} 
\partial^{\mu_{3}} \Pom^{\mu_{4}}_{\quad \nu_{2}}(x) -
\partial^{\mu_{4}} \Pom^{\mu_{3}}_{\quad \nu_{2}}(x) \big{)}        
\Big{]} \,.
\label{A5}
\end{eqnarray}
From (\ref{A4}) and (\ref{A5}) we get the bare vertex functions
$\Gamma_{\mu \nu,\kappa \lambda}'^{(\Pom f_{2 \Reg} M)}(q_{1}, q_{2})\mid_{\rm bare}$
and 
$\Gamma_{\mu \nu,\kappa \lambda}''^{(\Pom f_{2 \Reg} M)}(q_{1}, q_{2})\mid_{\rm bare}$, respectively.
The results are as in 
(\ref{vertex_pompomPS_11}) and (\ref{vertex_pompomPS_33})
but with the replacements
\begin{eqnarray}
g_{\Pom \Pom M}' & \to & g_{\Pom f_{2 \Reg} M}'\,, \nonumber \\
g_{\Pom \Pom M}'' & \to & g_{\Pom f_{2 \Reg} M}''\,.
\label{A6}
\end{eqnarray}
This concludes our discussion of the $\Pom f_{2 \Reg} M$ couplings.


\bibliography{refs}

\end{document}